\documentclass[12pt,titlepage]{article}
\usepackage[myheadings]{fullpage}
\usepackage{setspace} 
\usepackage{amsmath}
\usepackage{graphicx}
\usepackage[round,numbers,sort&compress]{natbib} 
\title{DNA entropic elasticity for short molecules
  attached to beads}

\author{Jinyu Li\\
  Department of Applied Mathematics,\\
  University of Colorado at Boulder, Boulder, CO
  \and Philip C. Nelson\\
  Department of Physics and Astronomy\\
  University of Pennsylvania, Philadelphia, PA
  \and M. D. Betterton\thanks{
           Corresponding author.  Address: 
           Department of Physics,
           University of Colorado at Boulder
           390 UCB,
           Boulder, CO~80309 U.S.A.,
           Tel.:~(303)735-6235, Fax:~(303)492-4066}\\
  Department of Physics,\\
  University of Colorado at Boulder, Boulder, CO }
\date{}
\begin{document}
\maketitle

\abstract{Single-molecule experiments in which force is applied to DNA
  or RNA molecules have enabled important discoveries of nucleic acid
  properties and nucleic acid-enzyme interactions. These experiments
  rely on a model of the polymer force-extension behavior to calibrate
  the experiments; typically the experiments use the worm-like chain
  (WLC) theory for double-stranded DNA and RNA. This theory agrees
  well with experiments for long molecules.  Recent single-molecule
  experiments have used shorter molecules, with contour lengths in the
  range of 1-10 persistence lengths.  Most WLC theory calculations to
  date have assumed infinite molecule lengths, and do not agree well
  with experiments on shorter chains.  Key physical effects that
  become important when shorter molecules are used include
  (\textit{i}) boundary conditions which constrain the allowed
  fluctuations at the ends of the molecule and (\textit{ii})
  rotational fluctuations of the bead to which the polymer is
  attached, which change the apparent extension of the molecule. We
  describe the finite worm-like chain (FWLC) theory, which takes into
  account these effects. We show the FWLC predictions diverge from the
  classic WLC solution for molecules with contour lengths a few times
  the persistence length.  Thus the FWLC will allow more accurate
  experimental calibration for relatively short molecules,
  facilitating future discoveries in single-molecule force microscopy.
  
  }


\section*{Introduction}

Single-molecule force microscopy, in which force is applied to DNA or
RNA molecules one at a time, has enabled important discoveries of the
mechanical behavior of nucleic acid (NA) molecules and NA-enzyme
interactions \citep{bustam00,nelson04}.  These experiments are usually
calibrated and interpreted using the worm-like chain theory (WLC). The
WLC predicts the average end-to-end extension $z$ of a semiflexible
polymer, given the force $F$ applied to the ends of the chain and the
values of two constant parameters.  The first parameter is the
molecule's unstressed total contour length $L$, which is proportional
to the number of base pairs in a double-stranded nucleic acid polymer.
The second parameter, the effective bending stiffness parameter or
persistence length $A$, describes the molecule's local elastic
behavior. Hence we expect the value of $A$ to depend on the molecule
type and the nature of the surrounding solution, but not on the
contour length. (Additional parameters enter when the stretching force
exceeds about 20 pN, or when the molecule is torsionally constrained.)
The WLC has been successfully applied to long molecules of DNA
\citep{bustam94,marko95} and RNA \citep{seol04,abels05} and it is now
the standard model used in calibrating and interpreting SM force
microscopy \citep{bouch99}.

Existing treatments of the WLC theory do not agree well with
experiments when the contour lengths of the NA molecules are short. In
particular, experiments on dsDNA molecules with $L/A \sim$ 1--10 yield
fit values of the persistence length which are physically unrealistic
\citep{liphar01,onoa03,perkin04}. These apparent values differ from the
accepted value, $A \approx 50$ nm, by factors of 2--5. However, the
persistence length is a material parameter, independent of the contour
length. Therefore this apparent $L$-dependent persistence length
amounts to a failure of the WLC, or at least its usual mathematical
treatment, which assumes that $L \gg A$.  In addition, the classic WLC
solution neglects some physical effects present in the experiments,
including the bead(s) attached to the end(s) of the molecule and
surface effects. In this paper we extend the WLC to include boundary
conditions at the ends of the polymer and rotational fluctuations of
the bead(s) attached to the molecule, and present solutions valid at
finite $L$.  Our ``finite worm-like chain'' (FWLC) solution gives more
accurate predictions than the classic WLC solution for polymers with
contour lengths less than a few times the persistence length. In
related recent work, Kulic \textit{et al.} have studied finite-length
effects in the stretching of DNA with a kink\citep{kulic05}; their
results apply in the high-force limit. At the other extreme,
references \citep{segal05} and \citep{nelson05} give numerical results
for the ``tethered particle'', or zero-force, limit.

\subsection*{Need for extensions of the WLC}

\begin{figure}
   \begin{center}
      \includegraphics*[width=5in]{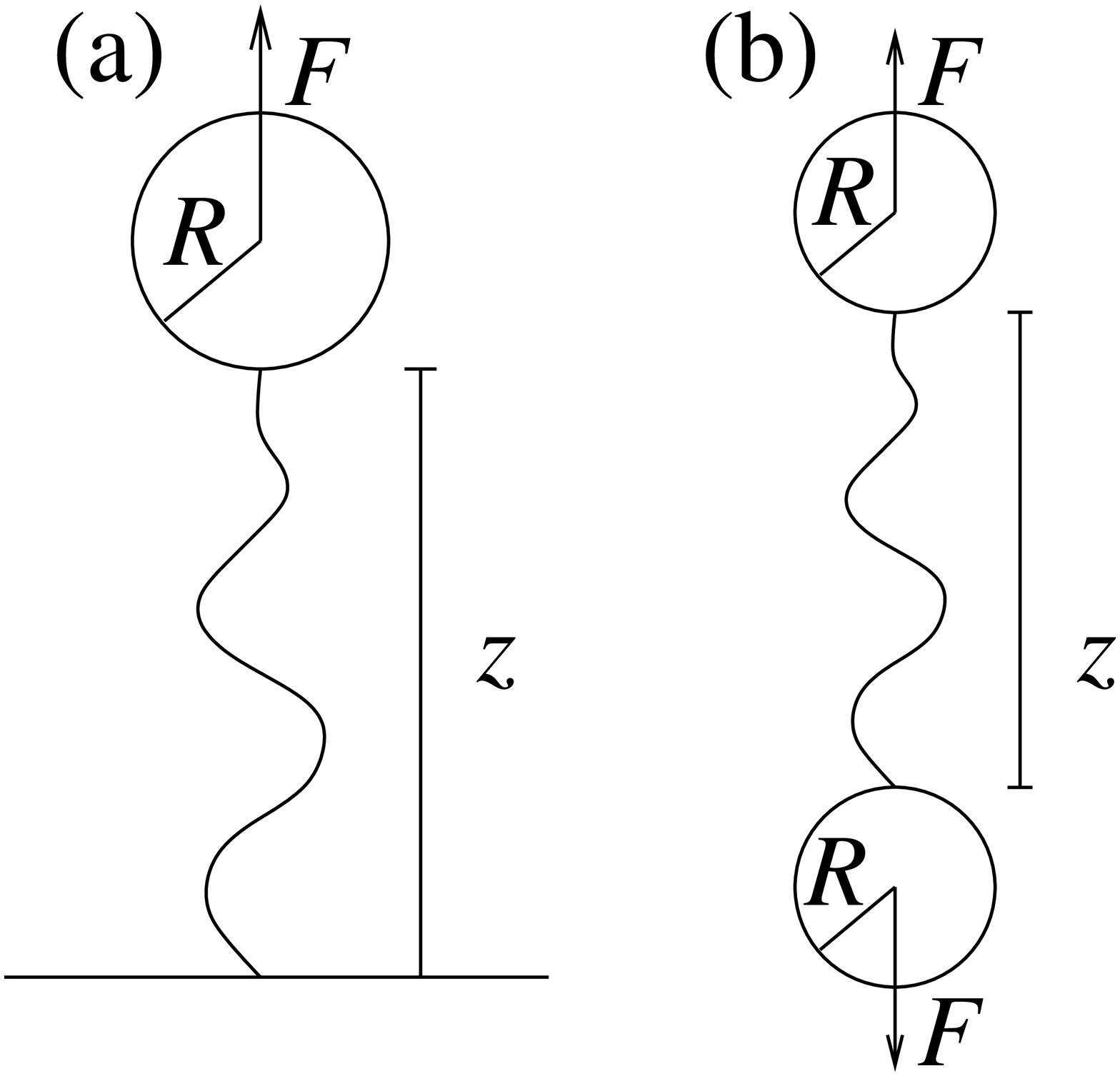}
      \caption{}
      \label{expgeom}
   \end{center}
\end{figure}

The interactions between proteins and nucleic acids are essential to
cells. Processes such as transcription and translation, splicing, DNA
copying, genome maintenance and DNA repair require NA-protein
interactions.  Premature aging, cancer, and even death can result from
defects in NA-protein interactions. (We use the abbreviation NA to
refer to single- or double-stranded DNA, RNA or DNA-RNA hybrid
molecules).  Single-molecule experiments have given new insight into
NA-protein interactions, yielding information that is difficult to
determine by other methods \citep{bustam00}.  In single-molecule force
microscopy, force is applied to assess NA mechanical behavior or
NA-protein interactions \citep{allem03}.  Typically one end of an NA
molecule is chemically attached to a surface and the other end is
attached to a bead (figure \ref{expgeom}).  Manipulation of the bead,
usually by optical \citep{neuman04} or magnetic \citep{gosse02} means,
stretches the NA polymer.  (In variants of this setup, the NA may be
attached to beads at both ends, or an NA-binding protein may be
attached to a surface.) Changes in the NA molecule extension and
force can be measured or applied.

A theory of NA stretching behavior is required to calibrate
single-molecule force experiments. The average end-to-end extension
$z$ of a polymer depends on the applied force $F$.  Using an accurate
model of the force-extension behavior means that once the calibration
is performed, knowledge of any two of the extension, force, and
contour length determines the other.  Thus, changes in experimentally
measured extension at fixed force can be directly related to changes
in contour length, for example when a processive enzyme ``reels in''
NA as it moves.

In the absence of an accurate model of NA force-extension behavior,
single-molecule force microscopy encounters several problems. First,
experimentalists lack a check on whether the trap calibration is
correct.  Second, without an accurate force-extension theory, the
contour length cannot be deduced from a measurement of the extension
at fixed force.  Therefore motion of motors which change the length of
the polymer in time cannot be accurately measured. Third,
force-extension measurements are frequently used to assess the number
of polymers attached to a given bead. The attachment of polymers to
the surface/bead occurs stochastically and can give more tethering
molecules than desired.  Measurement of the force-extension behavior
is used to confirm the correct number of tether molecules.  Fourth,
experiments which study forced unfolding of secondary structure
\citep{liphar01,liphar02,onoa03} require force-extension calibration to
determine the free energy of the folded molecule.

The worm-like chain theory (WLC) of NA elasticity has become the
standard model for calibrating experiments. The WLC assumes that the
polymer is an infinitely long, inextensible, isotropic rod with
bending rigidity.  The usual WLC solution gives the relationship
between fractional extension $z/L$ and force in the limit $L \to
\infty$ \citep{nelson04,marko95,bouch99}.  However, physical effects
neglected in the usual treatment of the WLC are important for
relatively short molecules, such as those used in several experiments
\citep{liphar01,liphar02,onoa03,perkin04,lang04}.

Experiments have fit data to the usual solution to the WLC even for
relatively short chains, because no alternative treatment has been
available \citep{onoa03,perkin04}. The result is a fit value of the
persistence length which depends on the contour length of the
molecule, with fit values as low as 10 nm \citep{onoa03}. For
double-stranded DNA, the accepted value of the persistence length is
approximately 50 nm (depending on solution conditions
\citep{wang97,bauman97}).  Experimental force-extension data, when fit
to a theory that correctly incorporates finite-length effects, should
recover the intrinsic value of $A$, independent of $L$.

\begin{figure}
   \begin{center}
      \includegraphics*[width=5in]{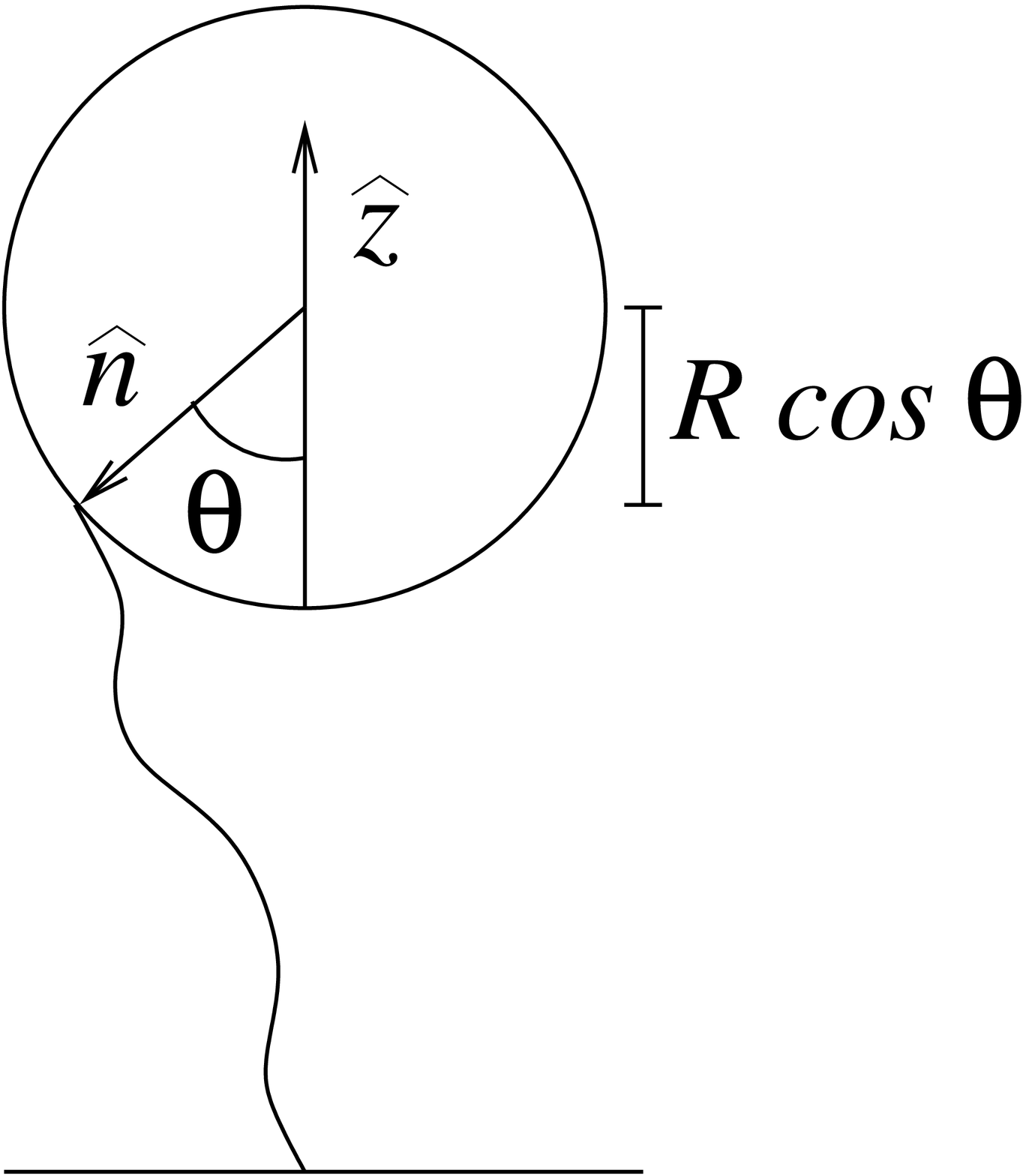}
      \caption{}
      \label{beadrot}
   \end{center}
\end{figure}

Finite-length effects can be incorporated into the WLC theory by
considering boundary conditions at the end of the chain
\citep{samuel02,kulic05}. The constraints imposed by the boundary
conditions alter the force-extension behavior.  A second important
physical effect typically neglected in the WLC theory is the
rotational motion of the bead(s) attached to the molecule (figure
\ref{beadrot}). Experiments measure the position of the center of the
bead, and typically one estimates the molecule extension by
subtracting the bead radius.  This estimate is correct only when the
polymer-bead attachment is in line with the direction of the force
($\theta=0$ in figure \ref{beadrot}) and the bead undergoes no
rotational fluctuations.  For long molecules (with contour lengths of
tens of $\mu$m), the fractional error introduced by subtracting the
bead radius (typically 50--500 nm) can be neglected. However, for
polymers with $L$ of few hundred nanometers, the error caused by
ignoring bead rotational fluctuations can be significant.

Several other physical effects are present in experiments, but
neglected in most treatments of the WLC. These include ({\it i}) the
exclusion interaction between the molecule and the surface (which
prevents the polymer from entering the solid surface), ({\it ii}) the
exclusion interaction between the bead and the surface, and ({\it
  iii}) the exclusion interaction between the molecule and the bead.
These interactions are most important for zero or very low applied
force, and have been addressed in recent work
\citep{segal05,bouch05,nelson05}.  Other interactions between the
surface and the chain or bead may occur, such as electrostatic or van
der Waals effects. In this paper, we will neglect these physical
effects, which are important only for very low forces or very long
polymers.  Our goal in this paper is to present a unified model which
includes the effects most important for single-molecule force
microscopy experiments on dsNA molecules with $L/A \sim$ 1--10; for
dsDNA this means contour lengths of 50--500 nm.

\section*{Theory}

The WLC model assumes an inextensible polymer: the contour length $L$
of the molecule (the total length of the molecular backbone) cannot be
changed by applied force.  For double-stranded DNA, the overstretching
transition which occurs for applied force around 60 pN is a dramatic
change in the organization of the molecule; however, lower applied
forces (up to approximately 20 pN) respect the inextensibility
constraint \citep{smith96,storm03}. The polymer is assumed to possess
an isotropic bending rigidity, characterized by the persistence
length, $A$. The persistence length is the length scale over which
thermal fluctuations randomize the chain orientation. (In principle,
the rod also resists twist.  However, in many experiments the twist is
unconstrained and can be ignored.)  We note that when considering
single finite-length molecules, we are not working in any
``thermodynamic limit''.  Therefore, different ensembles are not
guaranteed to be equivalent \citep{dhar02,keller03,sinha05}.  In this
paper we work in the ensemble relevant to most experiments, where the
applied force is fixed and calculate the extension.

\begin{figure}
   \begin{center}
      \includegraphics*[width=6in]{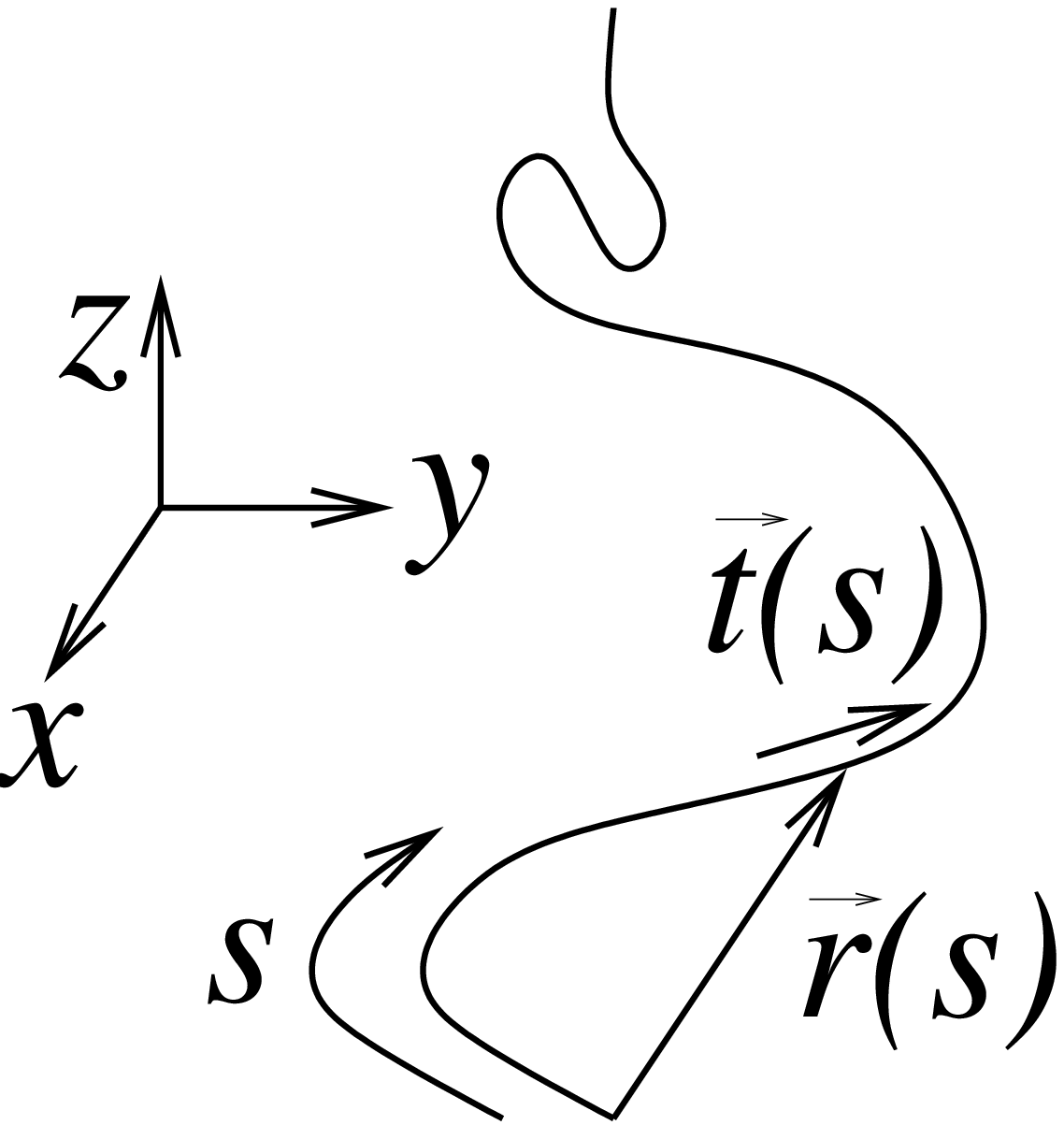}
      \caption{}
      \label{coord}
   \end{center}
\end{figure}

The chain energy includes the bending energy (Hooke's law in the chain
curvature) and the work done by the applied force.  The energy (in
units of the thermal energy $k_BT$) is
\begin{equation} 
E= - Fz + \int_0^L ds \ \frac{A}{2} \kappa^2
\end{equation}
Here $F$ (the force divided by the thermal energy $k_BT$) is applied
in the $\hat{\bf z}$ direction, $s$ denotes arc length, and the total
extension of the chain is $z$. The curvature can be defined in terms
of arc-length derivatives of the chain coordinate (figure
\ref{coord}). If the chain conformation is described by a space curve
${\bf r}(s)$ and the unit vector tangent to the chain is ${\bf
  \hat{t}}(s)$, then $\kappa = \left| \frac{\partial^2 {\bf
      r}}{\partial s^2}\right|= \left| \frac{\partial {\bf
      \hat{t}}}{\partial s}\right|$.  Note that the chain extension
can be calculated as $z = \int_0^L ds\ {\bf \hat{\bf z}}\cdot{\bf
  \hat{t}}$.

We rescale by dividing all lengths by the persistence length $A$, so
the scaled energy is
\begin{equation} 
E=\int_0^\ell ds' \left( \frac{\kappa'^2}{2} - f {\bf \hat{\bf z}}\cdot{\bf
      \hat{t}}  \right),
\end{equation}
where $\ell=L/A$, $f=F A$, $s'=s/A$, and $\kappa'=\kappa A$. We will
drop the primes in the remainder of the paper.

To determine the extension for a given applied force requires
averaging over different polymer conformations.  This leads to a path
integral formulation of the statistical-mechanics problem in the
tangent vector to the chain \citep{fixman73,yamak76}.  If the ends of
the chain are held at fixed orientations, the partition function of
the chain is
\begin{equation} 
Z=\int D{\bf \hat{t}} \ \exp \left[- \int_0^\ell ds \left( \frac{1
      }{2}(\partial_s{\bf \hat{t}})^2 - f {\bf \hat{\bf z}}\cdot{\bf
      \hat{t}} \right)\right],
\end{equation} 
where the integral in $D{\bf \hat{t}}$ is over all possible paths
between the two endpoints of the chain with the specified
orientations.  The partition function can be rewritten as a propagator,
which connects the probability distribution for the tangent vector at
point $s$, $\psi({\bf \hat{t}},s)$ to the same probability
distribution at point $s'$:
\begin{equation}
\psi({\bf \hat{t}},s) = \int d{\bf \hat{t}}' \ Z({\bf \hat{t}},s;{\bf
  \hat{t}}',s')\ \psi({\bf \hat{t}}',s').
\label{propagator} 
\end{equation}
From this relation, one can derive a Schr\"odinger-like equation which
describes the $s$ evolution of $\psi$ \citep{marko95}:
\begin{equation} 
\frac{\partial \psi}{\partial s}=\left(\frac{\nabla^2}{2}+f \cos
  \theta \right) \psi.
\label{hamilt}
\end{equation}
Here $\nabla^2$ is the two-dimensional Laplacian on the surface of the
unit sphere and $\cos \theta= \hat{\bf z} \cdot \hat{\bf t}$.

\subsection*{Additional physical effects included in the FWLC}

\subsubsection*{Boundary conditions at ends of chain}

For relatively short NA molecules, the boundary conditions at the ends
of the chain must be considered \citep{samuel02,kulic05}.  The boundary
conditions are specified by two probability density functions,
$\psi({\bf \hat{t}},s=0)$ and $\psi({\bf \hat{t}},s=L)$. These enter the
full partition function via
\begin{equation}
  \label{fullpart}
  Z_{tot}=\int d{\bf \hat{t}}_i \ d {\bf \hat{t}}_f \ \psi({\bf
    \hat{t}}_i,0) Z({\bf \hat{t}}_i,0;{\bf \hat{t}}_f,L) \psi({\bf
    \hat{t}}_f,L).
\end{equation}

\subsubsection*{Bead rotational fluctuations}

Rotational fluctuations of the bead can be explicitly included in the
theory. Here we describe the extension due to a bead at one end of the
chain, with the other end of the chain held at fixed orientation
(figure \ref{expgeom}a; the
generalization to consider beads at both ends of the polymer is
straightforward). In this case, the theory predicts the extension to
the center of the bead, not just the molecule extension.  We assume
that the bead is spherical and the fluctuations in the polymer-bead
attachment are azimuthally symmetric about the $\hat{\bf z}$ axis
(figure \ref{beadrot}).

The unit vector ${\bf \hat{n}}$ points from the center of the bead to
the polymer-bead attachment.  Thus the bead's contribution to the
energy of a fluctuation with a given ${\bf \hat{n}}$ is $F R {\bf
  \hat{n}} \cdot {\bf \hat{z}}$, where $R$ is the bead radius. Note
that the energy is minimized when ${\bf \hat{n}}=-\hat{\bf z}$, and
for other bead orientations the energy increases. This gives a
restoring torque on the bead which increases as $FR$ increases. We
write the rescaled energy including this term as
\begin{equation} 
E= f r {\bf
  \hat{n}} \cdot {\bf \hat{z}}+\int_0^\ell ds \left(
  \frac{\kappa^2}{2} - f \hat{{\bf z}}\cdot \hat{\bf t} \right),
\end{equation}
where $r=R/A$.  The partition function of the chain is
\begin{equation} 
Z=\int D{\bf \hat{t}} \ \exp\left[- \int_0^\ell ds \left( \frac{1
      }{2}(\partial_s{\bf \hat{t}})^2 - f {\bf \hat{z}}\cdot{\bf
      \hat{t}} \right)\right]
      \int_{\mbox{constr}} d {\bf \hat{n}} \exp \left(- f r {\bf
          \hat{n}} \cdot {\bf \hat{z}} \right),
\label{partfunbead}
\end{equation} 
where the first integral is over all possible paths between the two
endpoints of the chain, and the second integral is over all allowable
vectors ${\bf \hat{n}} $.  Below we evaluate the second integral,
thereby performing the average over fluctuations in ${\bf \hat{n}}$.


\section*{Methods}

Here we describe the calculation of the force-extension relation, the
main quantity of interest in single-molecule experiments.  We must
first calculate the tangent-vector probability distribution $\psi({\bf
  \hat{t}},s)$ for all $s$ along the chain. The distribution satisfies
equation \eqref{hamilt} above.  This PDE can be solved using
separation of variables in $s$ and ${\bf \hat{t}}$, where the angular
dependence is expanded in spherical harmonics \citep{marko95}.
\begin{equation} 
\psi({\bf \hat{t}},s) = \sum_{j=0}^{\infty} \Psi_j(s) Y_{j0}({\bf \hat{t}}).
\end{equation} 
We assume azimuthal symmetry, so only the $m=0$ spherical harmonics,
with no $\phi$ dependence, are included.  In the basis of spherical
harmonics, the operator in equation (\ref{hamilt}) is a symmetric
tridiagonal matrix $H$ with diagonal terms
\begin{equation} 
H_{j,j}=-\frac{j(j+1)}{2},
\end{equation} 
and off-diagonal terms
\begin{equation} 
H_{j,j+1}=\frac{f(j+1)}{(2j+1)(2j+3)}.
\end{equation}
The vector of coefficients at point $s$ is given by the matrix
exponential of $H$:
\begin{equation} 
{\bf \Psi}(s)=e^{sH}{\bf \Psi}(0).
\label{propagate}
\end{equation} 
This expression allows us to compute the probability distribution of
the tangent vector orientation at any point along the chain. This
result is exact if the infinite series of spherical harmonics is used.
In practice, the series must be truncated for numerical calculations.
Our calculations use $N=30$ unless otherwise specified.

The partition function is then calculated from the inner product
\begin{eqnarray} 
Z&=&{\bf \Psi}^T(s=\ell)e^{\ell H}{\bf \Psi}(s=0),\\
&=&\sum_{j,k}  \Psi_j(s=\ell) [e^{\ell H}]_{jk} \Psi_k(s=0).
\label{zinnprod}
\end{eqnarray} 
The extension at a given applied force is 
\begin{equation} 
\frac{z}{L}=\frac{1}{\ell}\frac{\partial \ln Z}{\partial f}.
\label{forcext}
\end{equation} 
This formula applies for a chain of any length.

\subsection*{Finite-length correction}

To calculate the force-extension relation we must determine $M=e^{\ell
  H}$ [equations \eqref{propagate} and \eqref{zinnprod}].  Because $H$
includes at least one positive eigenvalue, the entries $M$ grows
rapidly with $\ell$. This increase can lead to numerical overflow
errors when computing $M$.  However, we are interested not in the
entries of the matrix but in the logarithmic derivative of the
partition function, and a rescaling can avoid the overflow problem.

Let $A=e^H$ and denote by $\lambda_*$ the largest eigenvalue of $A$.
Then $M=A^{\ell}$.  If we define ${\cal A}=A/ \lambda_*$, we have $M =
\lambda_*^{\ell} {\cal A}^{\ell}$. Thus ${\cal A}$ has eigenvalues
with magnitude less than or equal to 1. The partition function can be
written
\begin{equation}
  Z=  \lambda_*^{\ell} {\bf \Psi}^T(s=\ell){\cal A}^{\ell} {\bf \Psi}(s=0).
\end{equation}
The logarithm of the partition function is then
\begin{equation}
  \ln Z=  \ell \ln \lambda_* + \ln [{\bf \Psi}^T(s=\ell){\cal A}^{\ell}
  {\bf \Psi}(s=0)]. 
\label{znew}
\end{equation}
In the usual WLC solution, only the first term is considered
\citep{marko95}, an approximation which is exact in the limit $\ell \gg
1$. The second term is the correction to $\ln Z$ due to finite-length
effects.

\subsection*{Boundary conditions}

The boundary conditions at $s=0$ and $s=\ell$ affect the
force-extension relation, because they affect the partition function
[equation \eqref{zinnprod}]. The functional form of a specific
boundary condition is given by the vector of coefficients ${\bf
  \Psi}(s=0)$, which is determined by the projection of $\psi(\hat{\bf
  t},s)$ onto spherical harmonics. To apply different boundary
conditions, we simply determine the partition function for different
vectors ${\bf \Psi}(s=0)$ and ${\bf \Psi}(s=\ell)$ \citep{samuel02}.

\begin{figure}
   \begin{center}
      \includegraphics*[width=5in]{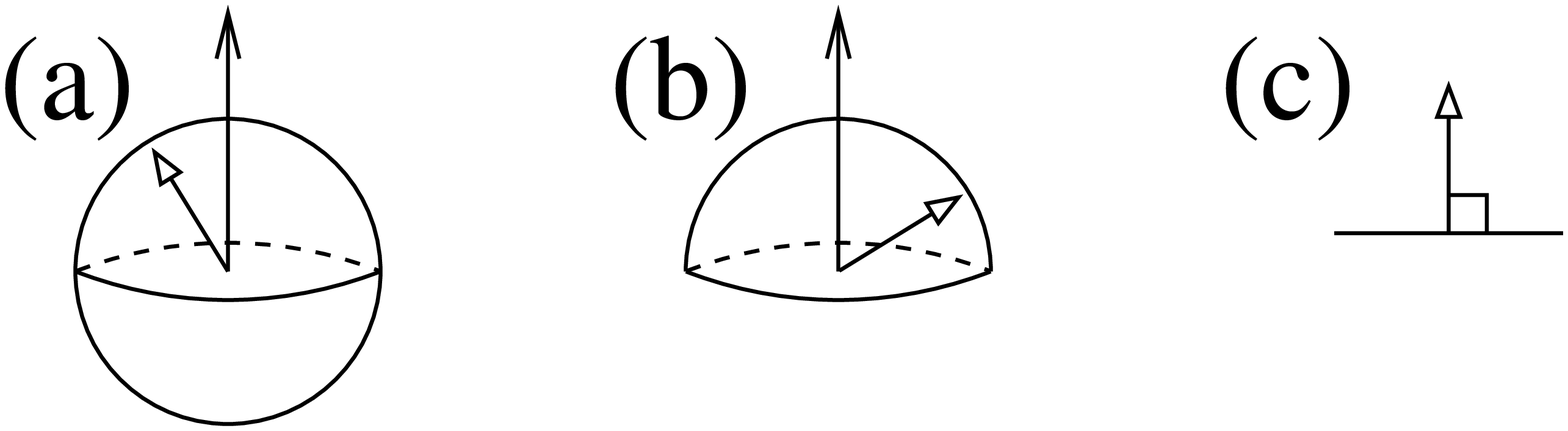}
      \caption{}
      \label{boundcond}
   \end{center}
\end{figure}

In this paper, we consider three types of boundary conditions (figure
\ref{boundcond}). First, in the ``unconstrained'' boundary condition
the tangent vector at the end of the chain is free to point in any
direction on the sphere (in $4 \pi$ of solid angle, figure
\ref{boundcond}a). In real experiments, however, the boundary
conditions are more constrained. If the polymer is attached to a
surface about a freely rotating attachment point, we might expect
``half-constrained'' boundary conditions (figure \ref{boundcond}b),
where the tangent vector at the end of the chain can point in any
direction on the hemisphere outside the impenetrable surface. Many
experiments appear to implement half-constrained boundary conditions,
because they use a flexible attachment between the chain and the
surface \citep{nelson05}. We also consider perpendicular boundary
conditions, where the tangent vector at the end of the chain is
parallel to the $\hat{\bf z}$ axis, normal to the surface (figure
\ref{boundcond}c).

For the unconstrained boundary condition, $\psi(\hat{\bf t})$ is
independent of $\cos \theta$. Therefore ${\bf \Psi}=(1,0,\cdots,0)$.
For the half-constrained boundary condition, the leading coefficients
of ${\bf \Psi}$ are (1, 0.8660, 0, -0.3307, 0, 0.2073, 0).  Finally, for
the perpendicular boundary condition the coefficients of ${\bf \Psi}$ are all
equal to 1. (Note that for the computation of the force-extension
relation, it is not necessary to properly normalize the probability
distribution, because we are computing the derivative of the logarithm
of $Z$.  Our expressions for the probability distribution vectors will
neglect the constant normalization factor.)

\subsection*{Bead rotational fluctuations}

The bead rotational fluctuations lead to an effective boundary
condition (at the end of the polymer) that depends on applied force
and bead radius.  For the single-bead experimental geometry (figure
\ref{expgeom}a), equation \eqref{partfunbead} gives the partition
function of the system including bead rotational fluctuations.  We can
perform the integral over ${\bf \hat{n}} $ to find 
the probability distribution of ${\bf \hat{t}} (\ell)$, that is, is
the effective boundary condition at the end of the polymer.  The
integral over ${\bf \hat{n}} $ is
\begin{equation} 
g({\bf \hat{t}}(\ell), {\bf \hat{z}}) =\int_{\mbox{constr}} d {\bf
  \hat{n}}\  e^{-f r {\bf 
          \hat{n}} \cdot {\bf \hat{z}}}.
\label{beadint}
\end{equation} 
Because the direction of $\hat{\bf n}$ is constrained relative to the
chain tangent $\hat{\bf t}(\ell)$, this integral is a function of
$\hat{\bf t}(\ell)$. By azimuthal symmetry, it depends only on the
scalar $\hat{\bf t}(\ell)\cdot \hat{\bf z}$. Accordingly we can
express equation \eqref{beadint} in the form $g({\bf \hat{t}}(\ell)
\cdot {\bf \hat{z}})$, where $g$ is the probability distribution of
tangent angles at $s=\ell$. (Note that we neglect the normalization
constant for $g$.) Expanding in spherical harmonics, we write
\begin{equation} 
g({\bf \hat{t}}(\ell)
      \cdot {\bf \hat{z}}) = g(\cos \gamma)=\sum_{j=0}^{\infty} \Psi_j(s=\ell)
      Y_{j0}(\gamma).
\label{bcbead}
\end{equation} 
The effective boundary condition depends on both the applied force and
the radius of the bead. The physical character of the chain-bead
attachment determines the constraints in the integral of equation
\eqref{beadint}, and therefore controls the $\Psi_j$.

For the two-bead experimental geometry, the partition function
includes integrals over bead rotational fluctuations for both ends of
the chain, and the effective boundary condition applies at both ends
of the polymer.

\subsubsection*{Half-constrained boundary conditions}

\begin{figure}
   \begin{center}
      \includegraphics*[width=3in]{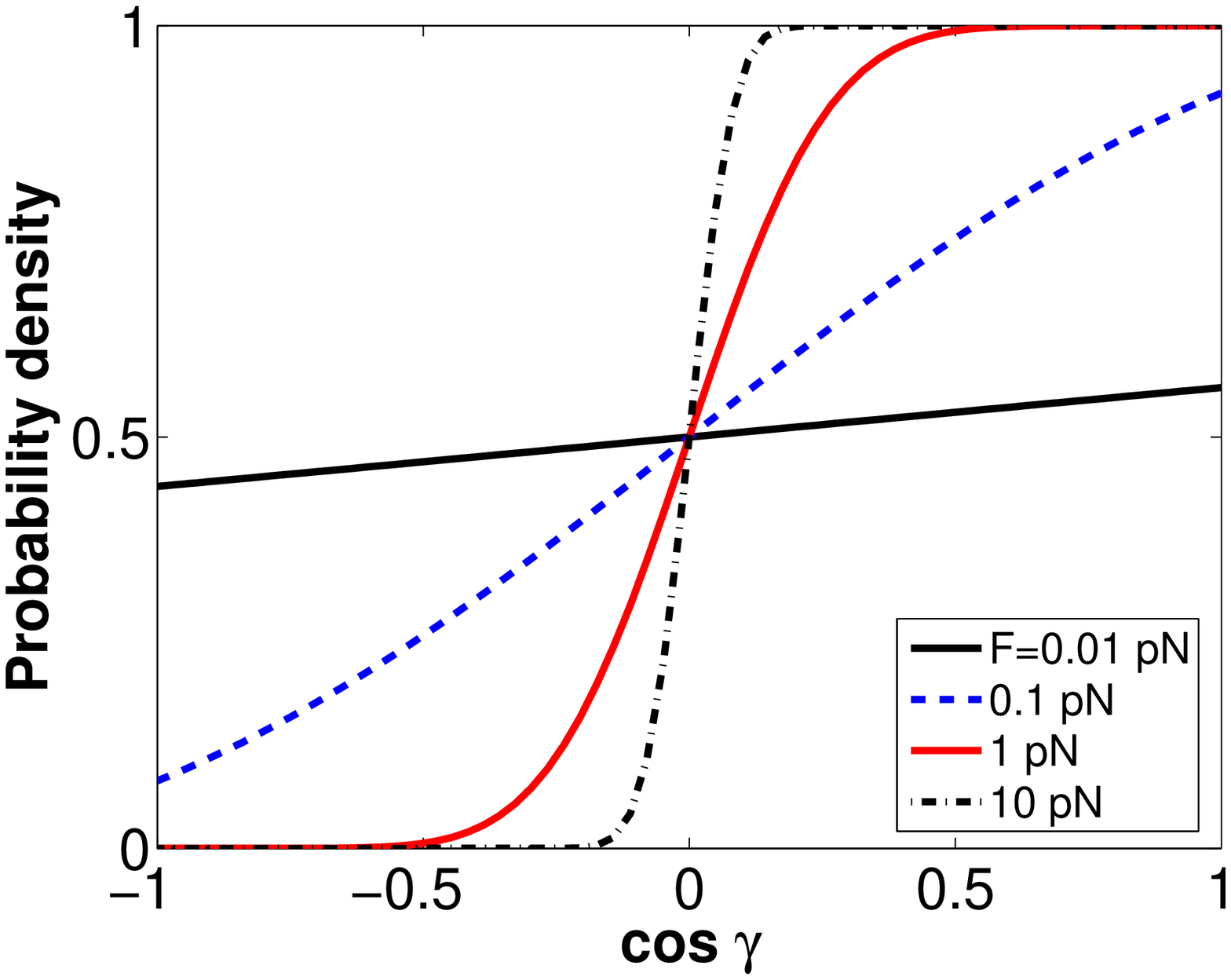}\\
      \includegraphics*[width=3in]{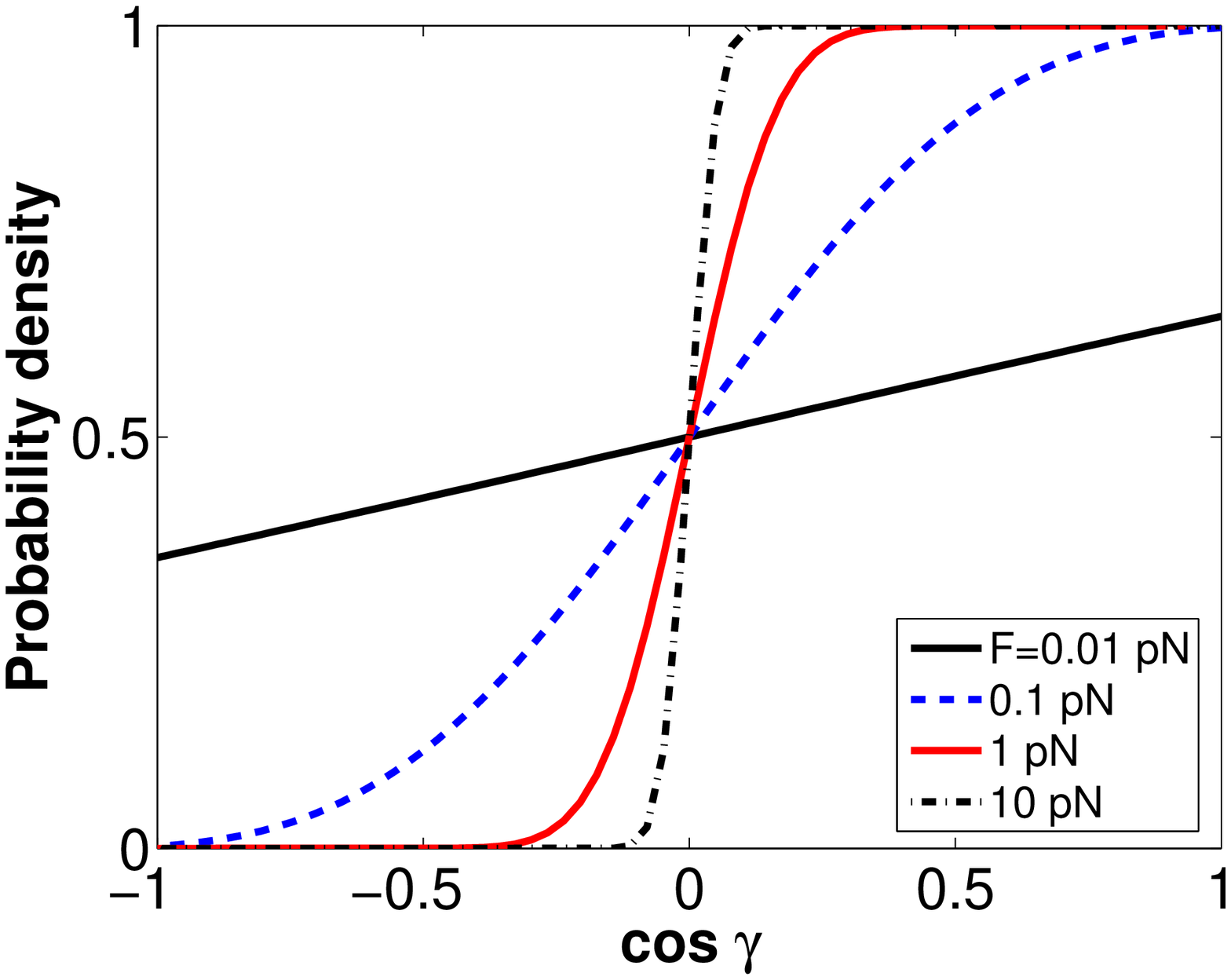}\\
      \includegraphics*[width=3in]{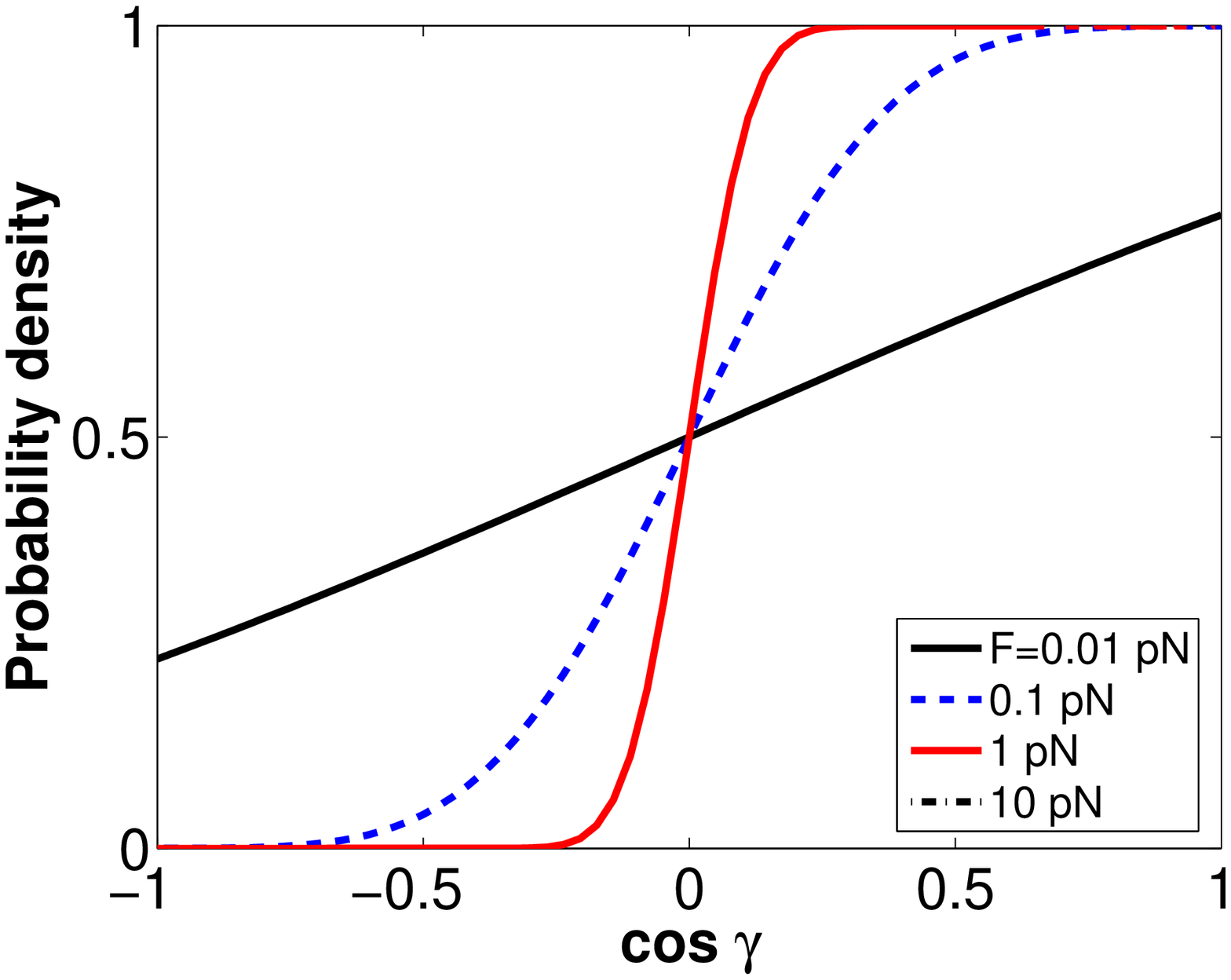}
      \caption{}
      \label{beadbc}
   \end{center}
\end{figure}

Suppose that the polymer-bead attachment is half-constrained, so the
tangent vector is free in a hemisphere (figure \ref{boundcond}b).
Then the integral of equation \eqref{beadint} is constrained by ${\bf
  \hat{t}}(\ell) \cdot {\bf \hat{n}}<0$, or
\begin{equation} 
g({\bf \hat{t}} \cdot {\bf \hat{z}} )=\int_{{\bf
\hat{t}} \cdot {\bf \hat{n}}<0} d {\bf \hat{n}} \ e^{-  f r {\bf
          \hat{n}} \cdot {\bf \hat{z}}}.
\end{equation} 
To evalute the integral, we choose a polar coordinate system where
${\bf \hat{t}} \cdot {\bf \hat{z}} =\cos \gamma$, ${\bf \hat{t}} $
points along the polar axis, and $\phi=0$ corresponds to the ${\bf
  \hat{z}} $ direction. Therefore ${\bf \hat{z}} = (\sin \gamma, 0,
\cos \gamma)$, ${\bf \hat{n}} = (\sin \theta \cos \phi, \sin \theta
\sin \phi, \cos \theta)$, and ${\bf \hat{n}} \cdot {\bf \hat{z}} =
\sin \theta \cos \phi \sin \gamma + \cos \theta \cos \gamma$. We can
then write $g$ in terms of the Bessel function $J_o$.
\begin{eqnarray} 
  g({\bf \hat{t}} \cdot {\bf \hat{z}} )&=&\int_0^{2 \pi} d \phi \int_{-1}^0 d
  \cos \theta \  \  e^{-  f r (\sin \theta \cos \phi \sin
    \gamma +   \cos \theta \cos \gamma)}, \\
  &=&\int_{-1}^0 d
  \cos \theta \  e^{-  f r \cos \theta \cos \gamma} \int_0^{2 \pi} d
  \phi  \  e^{-  f r \sin \theta \cos \phi \sin 
    \gamma },\\
  &=& 2 \pi \int_{-1}^0 d
  \cos \theta \  e^{-  f r \cos \theta \cos \gamma}
  J_o(-i f r \sin \theta \sin \gamma).
\label{gint}
\end{eqnarray} 
Note that $\int_0^{\pi} \exp (z \cos x)= \pi J_o(i z)$, and the
integral from $0$ to $2 \pi$ is even about $\pi$.  We show the values
of this function in figure \ref{beadbc} for different applied forces
and bead radii. As expected, for small applied force and small bead
radius, the probability distribution approaches a constant value
(independent of $\cos \gamma$). However, for large applied force, the
probability distribution approaches the half-constrained distribution
one would expect in the absence of bead rotational fluctuations.

\subsubsection*{Numerical methods}

Because $\cos \theta$ is integrated over negative values in equation
\eqref{gint}, the term $e^{- f r \cos \theta \cos \gamma}$ diverges as
$f r$ increases.  However, we only calculate the dependence of $g$ on
$\cos \gamma$ (correct normalization is not required). Therefore we
replace the term $e^{- f r \cos \theta \cos \gamma}$ in the integral
with the term $e^{- f r (\cos \theta \cos \gamma +1)}$.

To determine the effective boundary condition due to the rotationally
fluctuating bead requires that we expand integrals of the form given
in equation \eqref{gint} in spherical harmonics.  Direct numerical
projection of the integral onto spherical harmonics leads to large
errors, because numerical integration of rapidly oscillating functions
(such as the higher-order spherical harmonics) is inaccurate.  To
solve this problem, we used an interpolating basis (reference
 \citep{alper02}, section 3.1.4). This allows the coefficients of the
spherical harmonics to be determined by evaluation of equation
\eqref{gint} at specific points (corresponding to the zeros of the
Legendre polynomials).  The integral in equation \eqref{gint} was
evaluated numerically using Gauss-Legendre quadrature.

\section*{Results}

Here we delineate the regimes where the usual WLC solution applies,
and where the FWLC is required for good agreement with experiment.
The classic WLC predictions becomes less accurate as the contour
length of the chain decreases.  The differences between the WLC and
FWLC also depend on applied force, boundary conditions, and bead
radius.

\begin{figure}
   \begin{center}%
      \includegraphics*[width=3in]{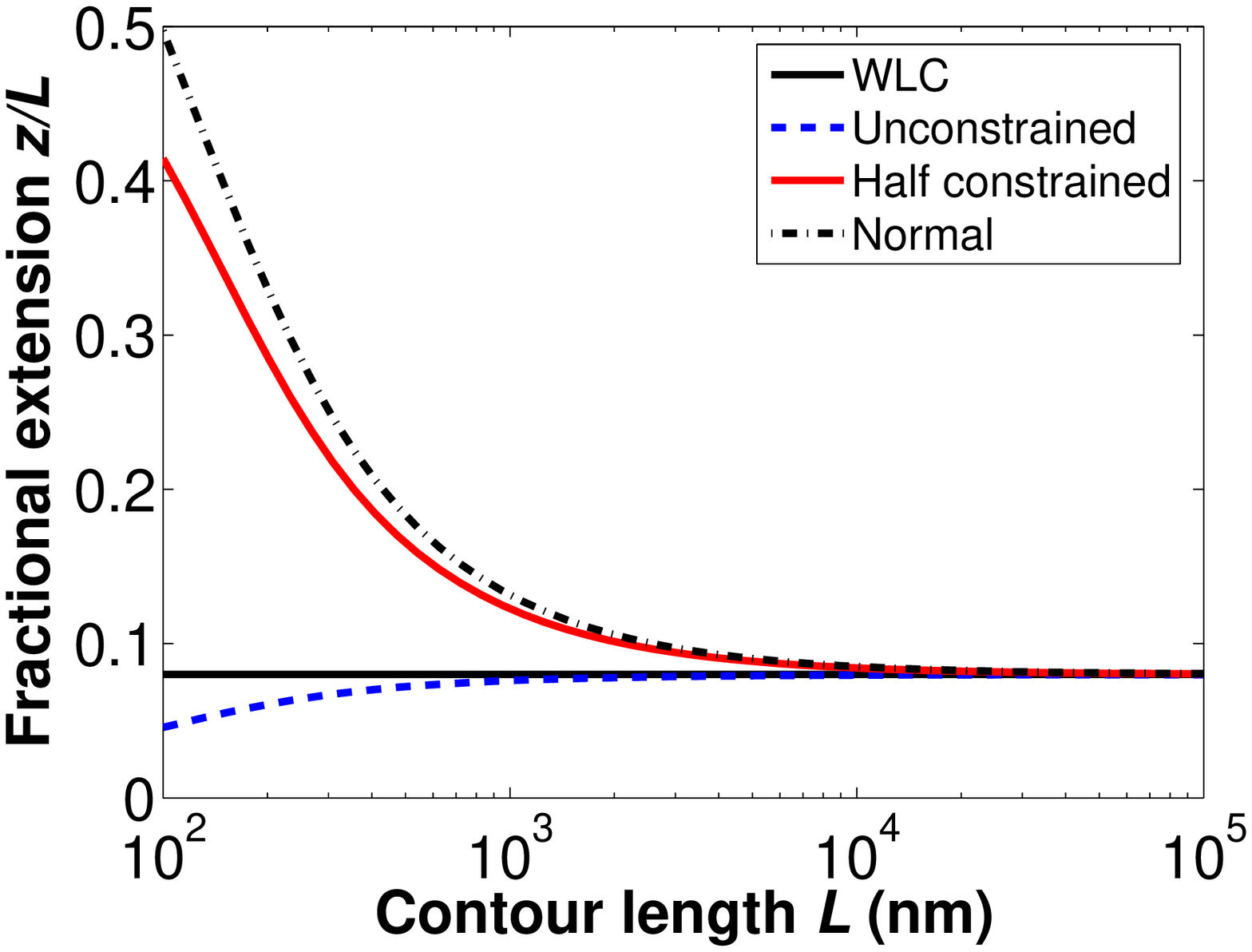}\\
      \includegraphics*[width=3in]{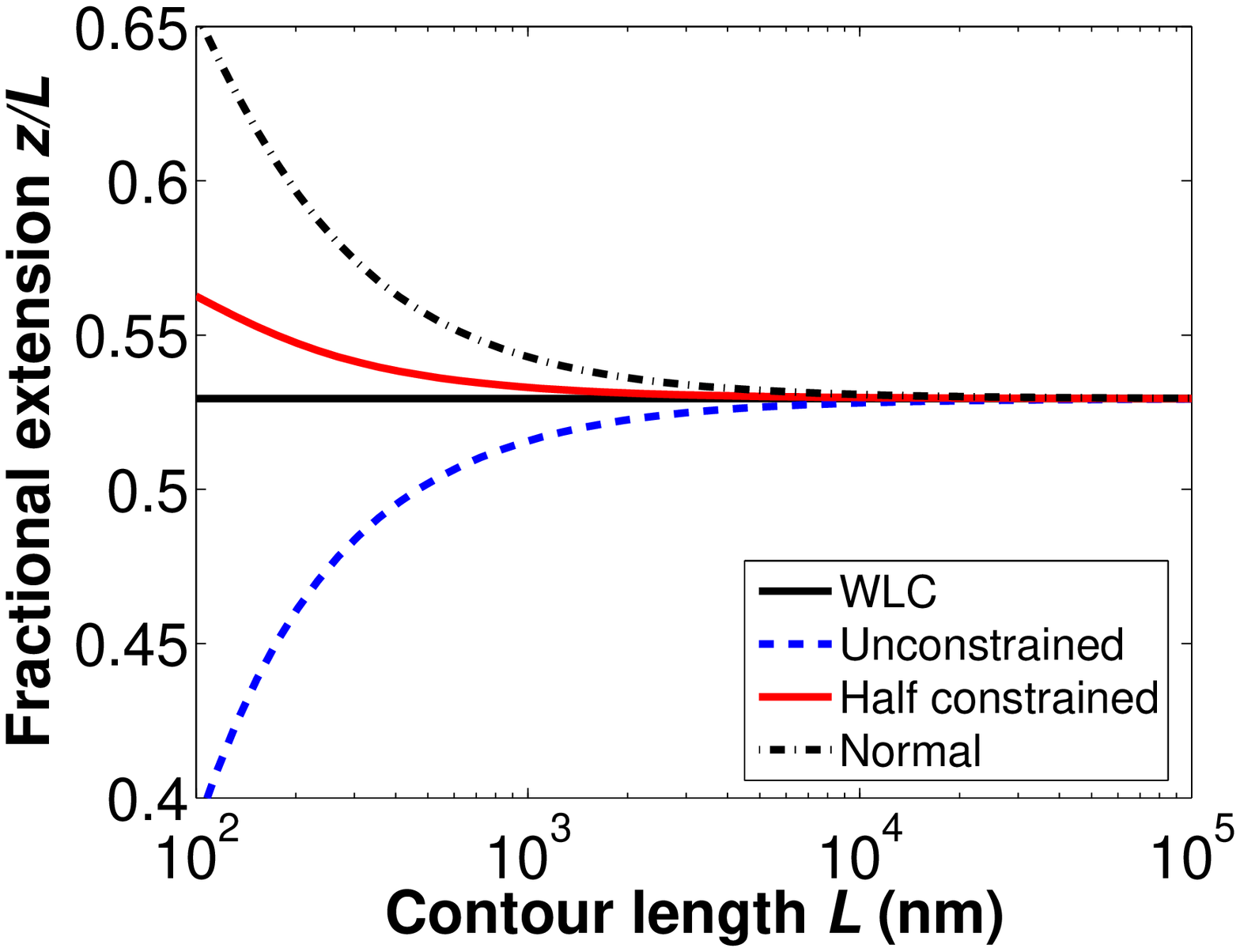}\\
      \includegraphics*[width=3in]{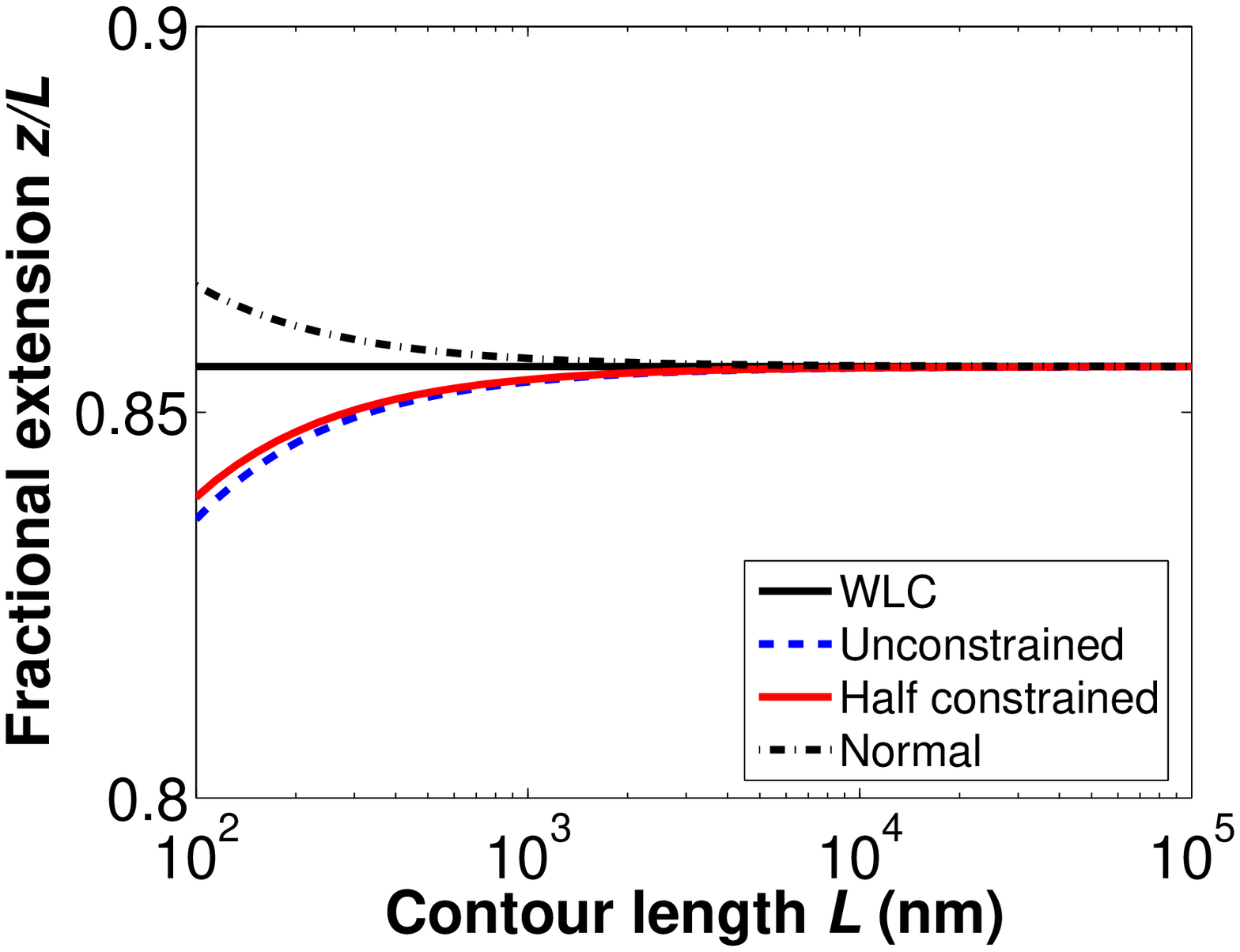}
      \caption{}
      \label{infconv}
   \end{center}
\end{figure}

As expected, the FWLC predictions converge to the WLC predictions as
the polymer contour length increases.  Figure \ref{infconv} shows the
predicted fractional extension $z/L$ as a function of the chain
contour length $L$ (for fixed applied force).  Predictions calculated
from the classic WLC solution are independent of contour length.
However, for the FWLC the predicted fractional extension deviates from
the infinite value for smaller values of $L$. In all cases, the FWLC
predictions converge to the classic WLC results as $L \to \infty$.


\subsection*{Error threshold contour length}

To summarize the different predictions of the FWLC and WLC, we
calculate the ``error threshold contour length'' $L_e$. As shown in
figure \ref{infconv}, the FWLC and WLC predictions diverge as $L$ is
decreased. We decrease $L$ and compare the FWLC and WLC predictions.
The error threshold contour length $L_e$ is defined as the contour
length where the relative difference between the two predictions first
reaches 5\%.  For contour lengths $L < L_e$ the FWLC model is
necessary for accurate predictions. By characterizing the dependence
of $L_e$ on boundary conditions, applied force, and bead radius, we
can understand which parameters control the difference between FWLC
and WLC predictions.

\begin{figure}
   \begin{center}
    \includegraphics*[width=3in]{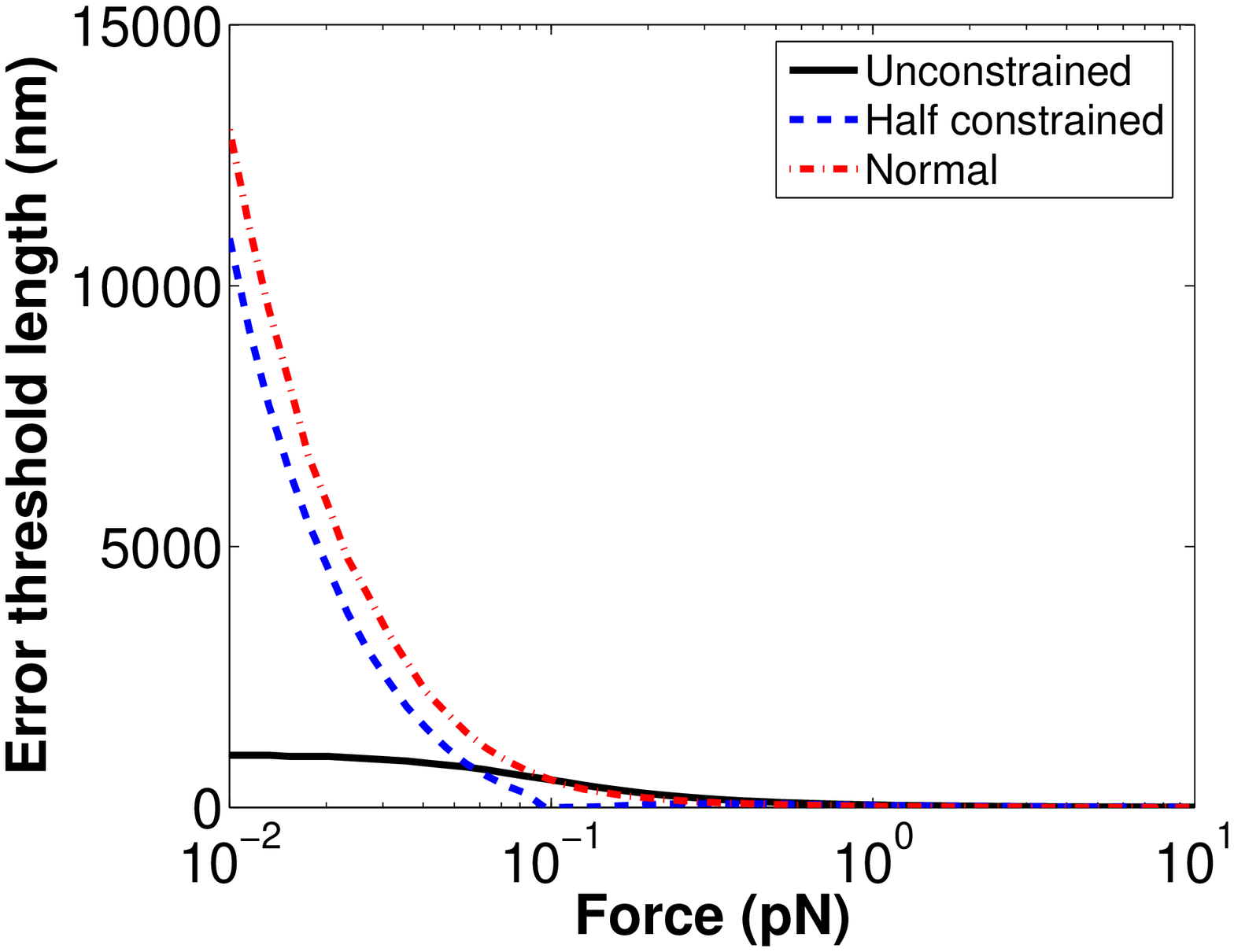}\\
      \includegraphics*[width=3in]{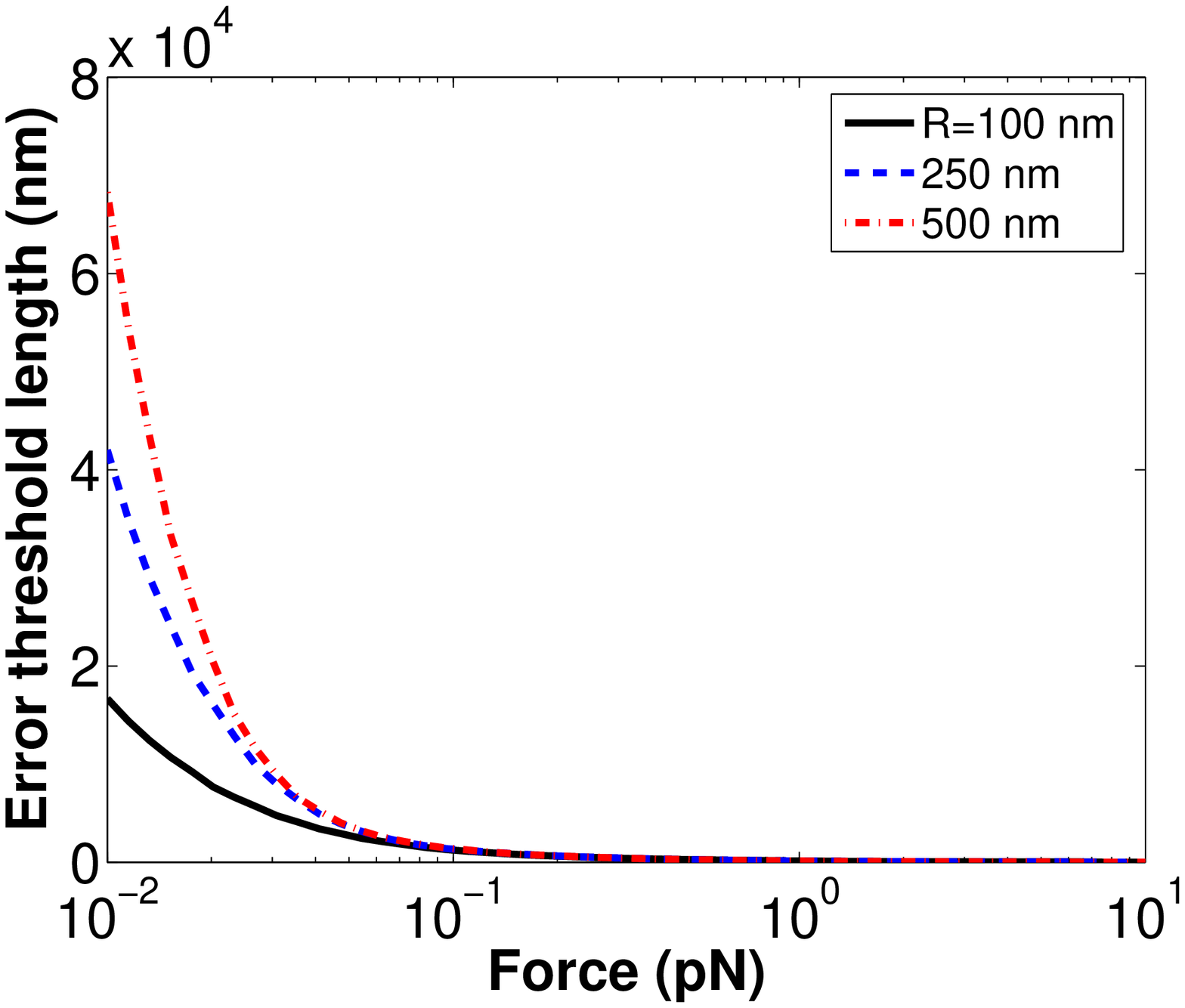}\\
      \includegraphics*[width=3in]{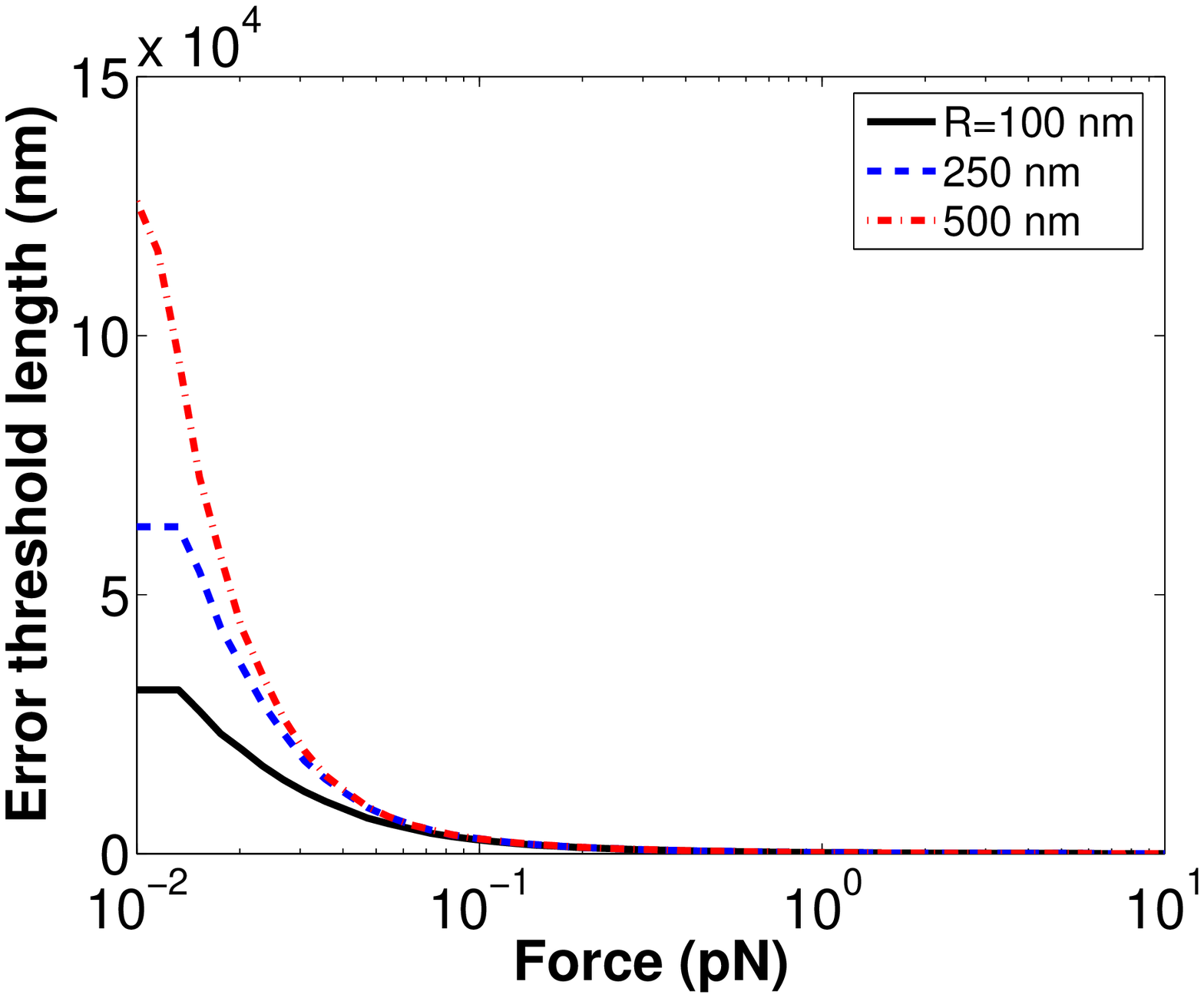}      
\caption{}
      \label{errthresh}
   \end{center}
\end{figure}

Figure \ref{errthresh} shows the error threshold length as a function
of applied force for different boundary conditions and bead radii.
For all conditions described, $L_e$ is largest for low applied force,
where the predicted extension is most altered by finite-length
effects. This behavior is intuitively reasonable.  When the applied
force tends to infinity, only one chain conformation is possible---the
chain is perfectly straight, aligned in the direction of the force. In
this case, modifying the theory to include boundary conditions or bead
rotational fluctuations does not alter the polymer conformation.  The
characteristic propagation length of deformations in a classical
elastic rod (that is, neglecting thermal fluctuations) is
$\sqrt{A/F}$. As the force increases, the effects of any constraints
on the ends of the polymer have decreasing effect on the conformation.
Similarly, when the applied force is large bead rotational
fluctuations require a large amount of energy and are suppressed.

When the applied force is low, many different chain conformations are
probable. Therefore the boundary conditions and bead rotational
fluctuations exert greater influence on the chain extension.
Unconstrained boundary conditions maximize the number of allowable
conformations. In this case the entropy is maximized, and the polymer
extension is low: increasing the extension requires excluding many
chain conformations, which requires more work. More constrained
boundary conditions restrict the number of conformations that are
possible, leading to lower entropy and therefore a larger extension.
At low force, the most constrained boundary conditions typically lead
to the largest extension.  We illustrate this idea with the limit of a
perfectly rigid rod which is constrained at one end to be perpendicular to
the surface.  The infinite rigidity means that the molecule must be
perfectly straight.  The boundary conditions require that the polymer
take only one conformation, perpendicular to the surface. In other words, the
boundary conditions severely restrict the number of available
conformations, and thereby increase the predicted extension.  (If the
boundary conditions are unconstrained, the rod rotates to point in
many different directions and its average extension is lower).  In
addition, for low applied force bead rotational fluctuations are
larger, leading to a larger alteration of the extension due to bead
fluctuations.

\subsection*{Force-extension behavior}

\begin{figure}
   \begin{center}
      \includegraphics*[width=3in]{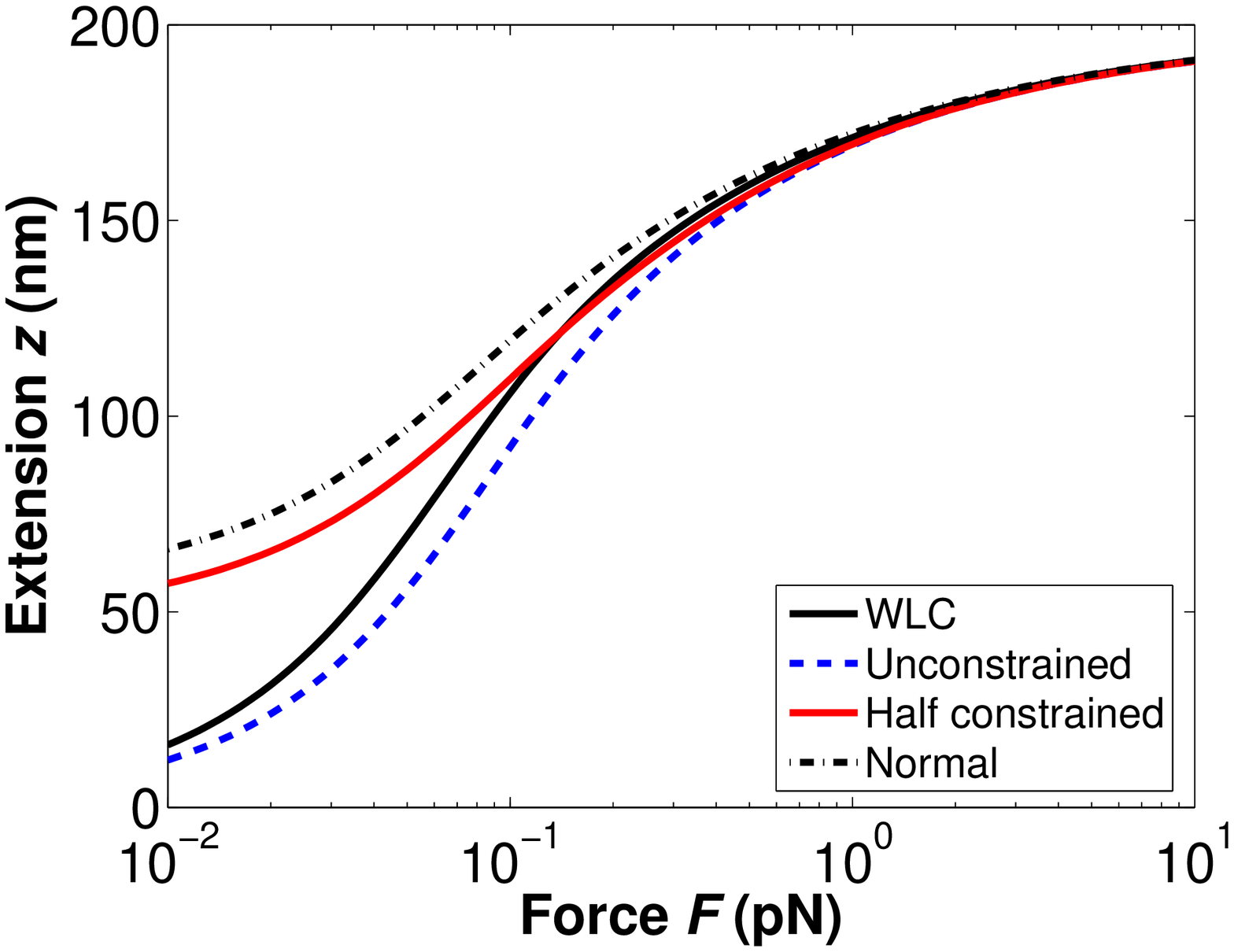}\\
      \includegraphics*[width=3in]{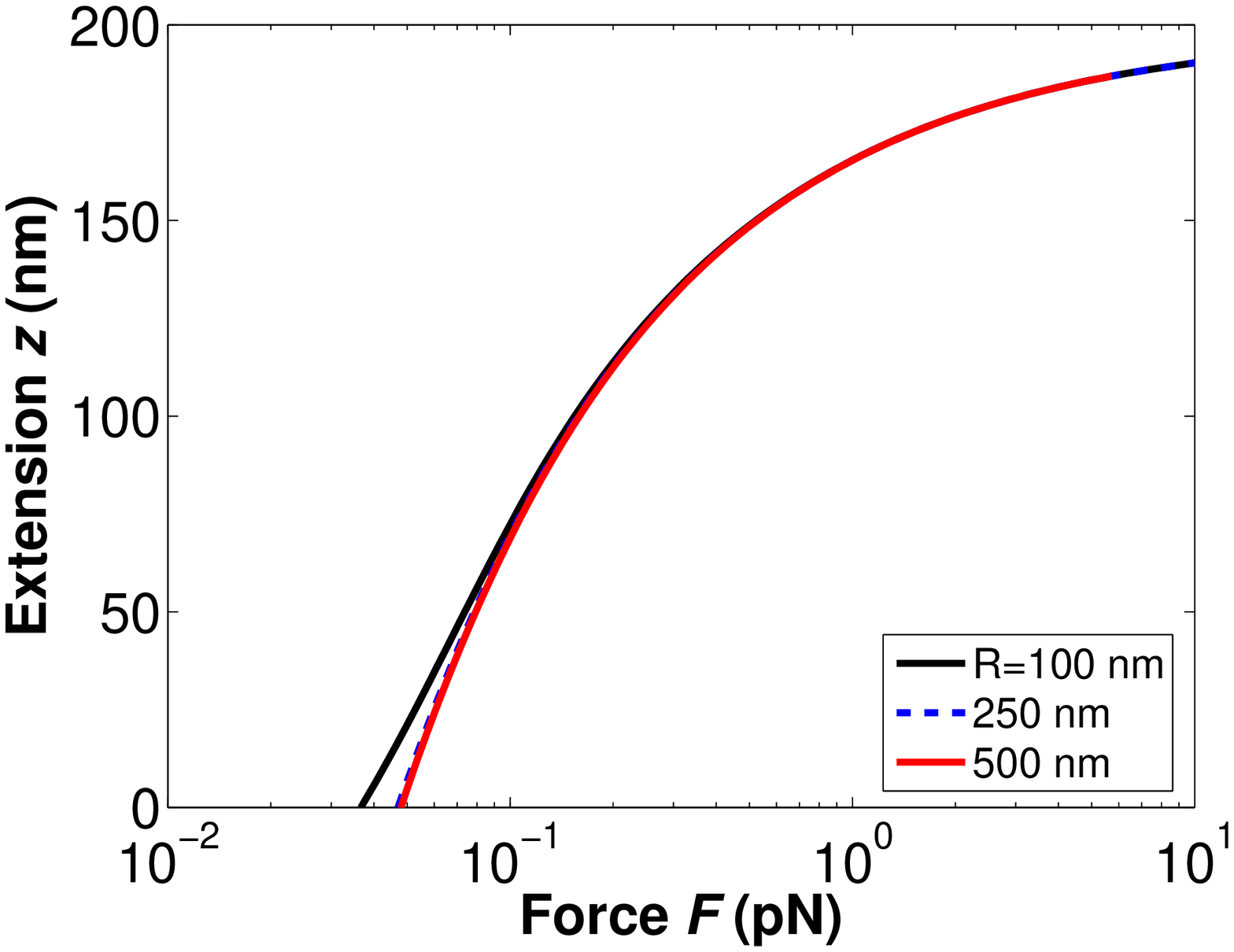}\\
      \includegraphics*[width=3in]{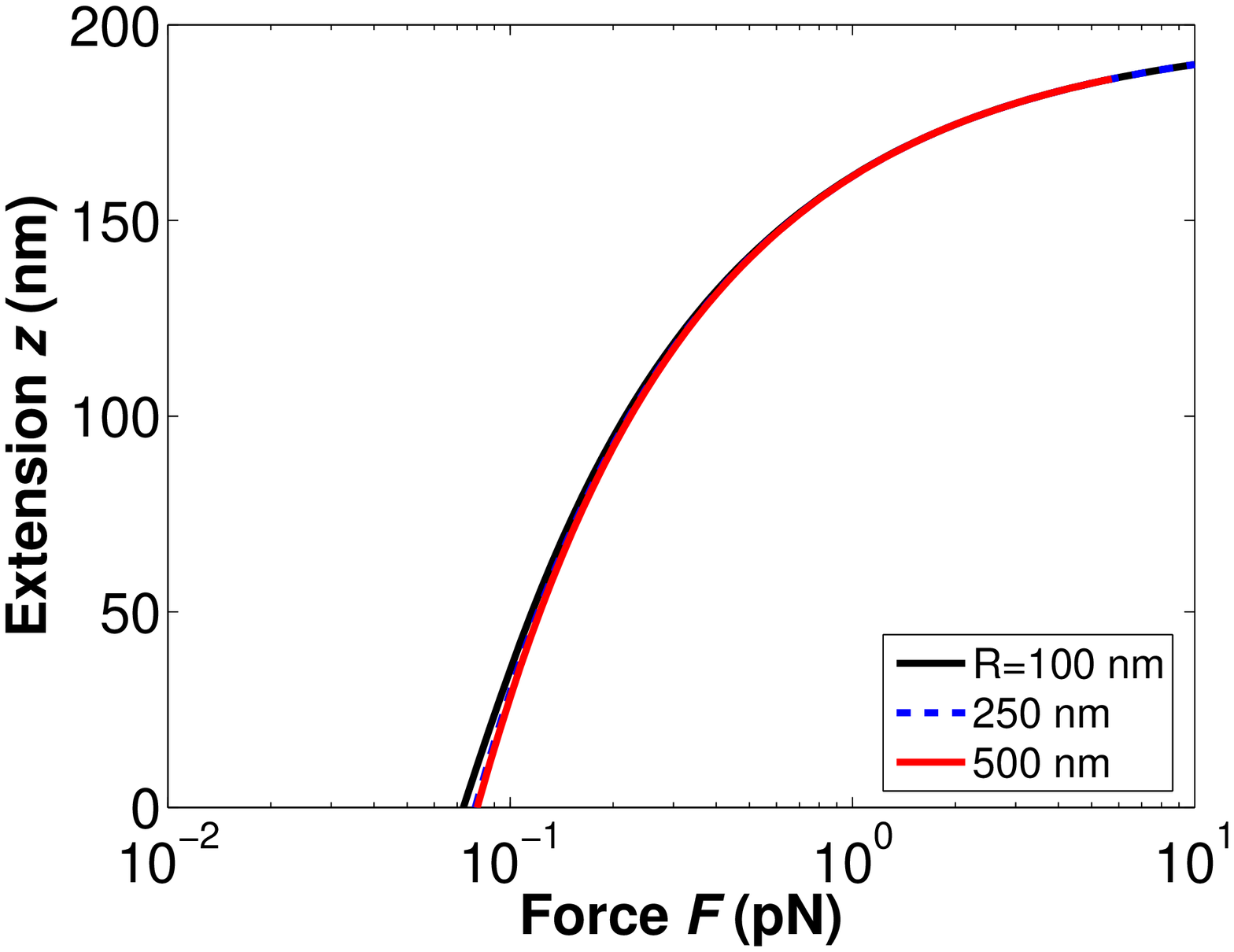}   
      \caption{}
      \label{forcextfig1}
   \end{center}
\end{figure}

\begin{figure}
   \begin{center}
      \includegraphics*[width=3in]{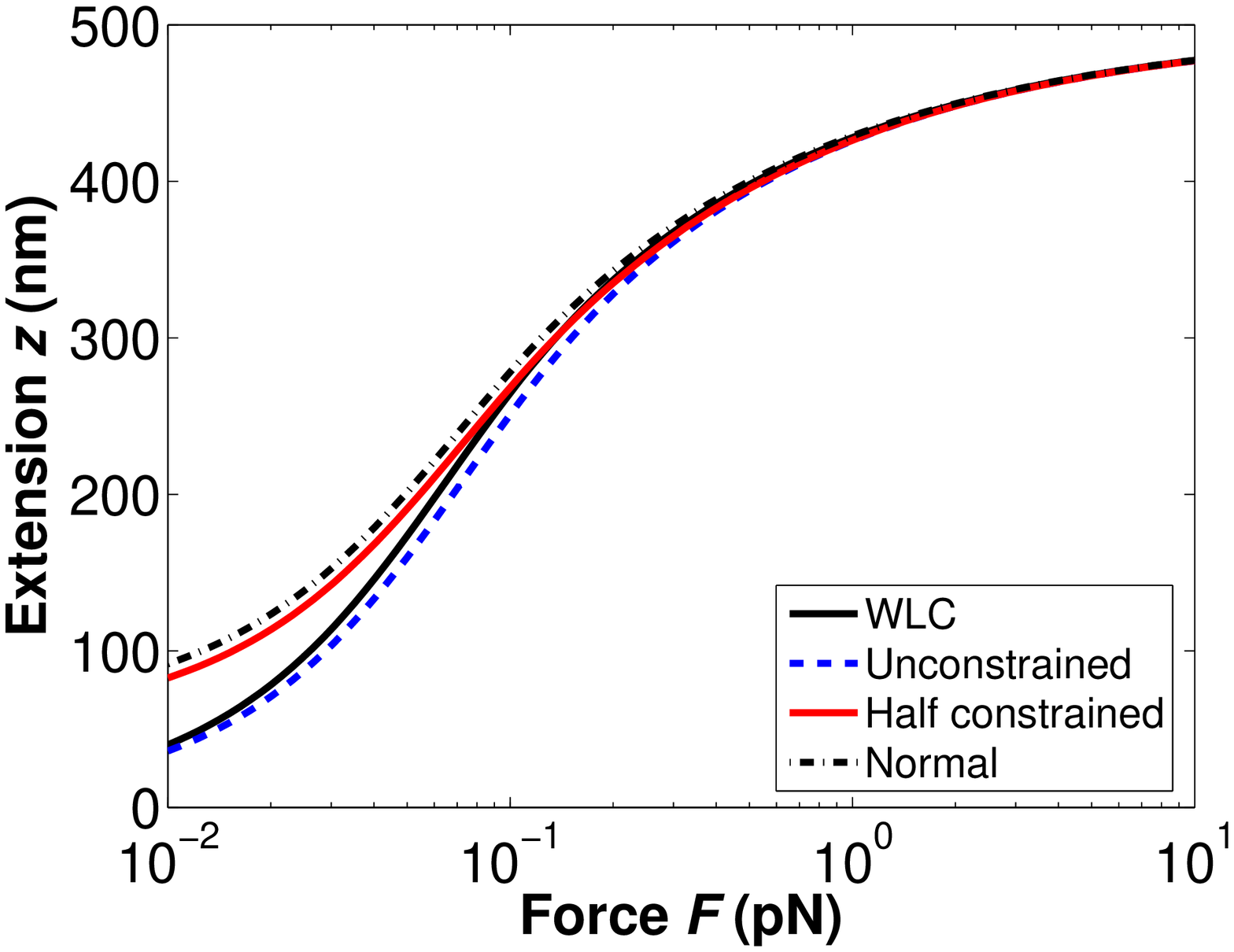}\\
      \includegraphics*[width=3in]{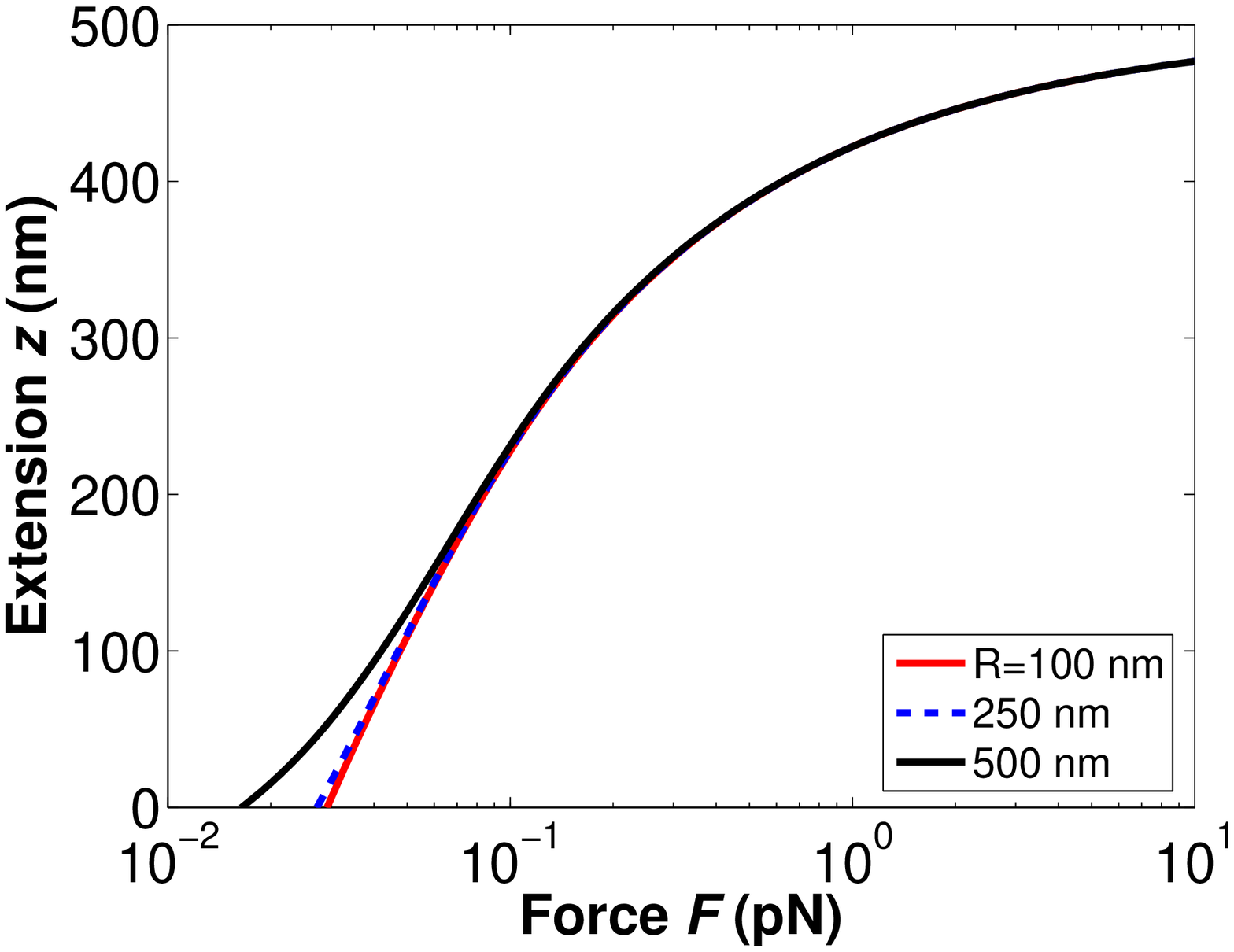}\\
      \includegraphics*[width=3in]{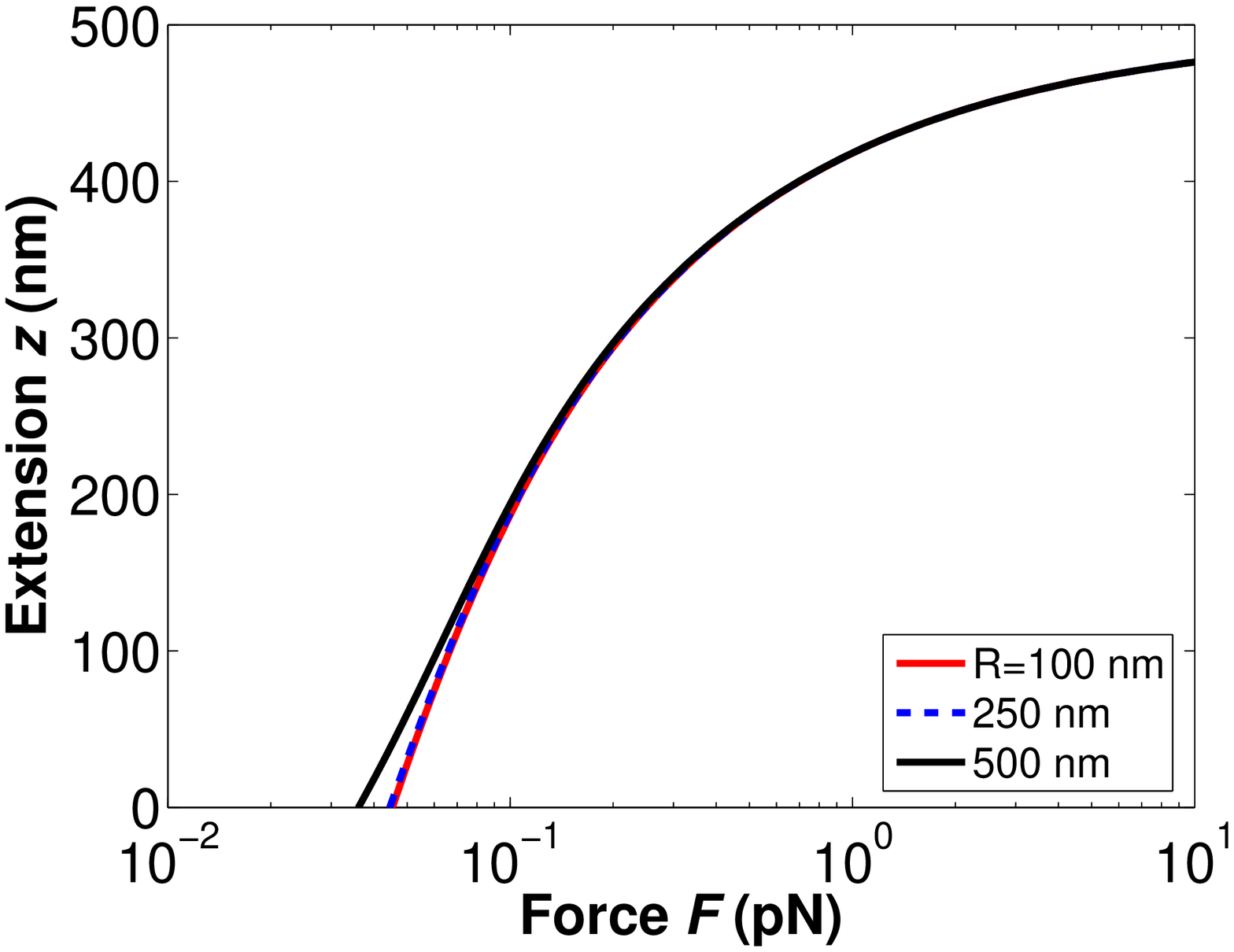}     
      \caption{}
      \label{forcextfig2}
   \end{center}
\end{figure}

We illustrate the force-extension curves for different boundary
conditions and contour lengths: $L/A=4$ in figure \ref{forcextfig1}
and $L/A=10$ in figure \ref{forcextfig2}. As described above, ({\it
  i}) the fractional difference between the infinite and finite
predictions is largest for low contour lengths and low applied forces,
({\it ii}) the more constrained boundary conditions lead to the
largest changes relative to the infinite model, and ({\it iii}) the
more constrained boundary conditions typically lead to larger
extensions at low force.  Therefore, the force-extension curves for
the FWLC with unconstrained boundary conditions are closest to the
classic WLC predictions (figure \ref{forcextfig1}). 

Including bead rotational fluctuations in the model leads to an
effective boundary condition that depends on applied force and bead
radius, as shown in figure \ref{beadbc}. The variation of the
effective boundary condition with force alters the shape of the
force-extension curve. For high applied force, the fluctuations
require more energy and so large-angle fluctuations occur less
frequently.  Therefore, in the limit $F R \to \infty$ the probability
distribution approaches that of the chain-bead attachment.  However,
for smaller $F R$, the bead rotational fluctuations ``smear out'' the
probability distribution, making the effective tangent angle boundary
condition approach a constant, independent of $\cos \gamma$.  For
smaller $F R$, the the predicted extension is similar to that expected
for no bead and unconstrained boundary conditions. This typically
leads to a smaller extension than would occur in the absence of a
bead. For larger $F R$, the effective boundary condition approaches
the boundary condition that would occur with no bead. In this case,
there is little difference between the predicted extension with and
without a bead.

We note that two opposing trends occur as the bead radius varies: when
the bead is smaller, the fractional error made by subtracting $R$ from
the measured extension to estimate the molecule extension is smaller,
simply because one is subtracting a smaller value. However, for
smaller $R$ the restoring torque on the bead for a given $F$ is also
smaller, which leads to larger angular fluctuations for smaller beads.

At very low forces our FWLC theory gives inaccurate predictions, because
it does not explicitly include the exclusion interaction between the
bead and the wall. If the force is very low, the predicted extension
would imply that the bead overlaps with the wall (or the other bead).
Typically we find that applied force $>0.08$ pN is required to prevent
predicted bead-wall overlap in our theory.  Although the FWLC does
lead to unphysical predictions at very low forces, the simple model of
the bead rotational fluctuations is useful for larger values of the
force.

\subsection*{Fitting FWLC curves to the classic WLC solution}

Here we illustrate the typical errors that occur if one fits
force-extension curves generated by the FWLC to the usual WLC
solution.  We generated force-extension data with the FWLC, and fit to
the WLC solution of Bouchiat \textit{et al} \citep{bouch99}. We find
that the contour length is typically fit well by the WLC model, but
large errors in the fitted value of the persistence length occur. The
apparent persistence length decreases with contour length (figure
\ref{fitper}), as observed experimentally in the stretching of short
DNA molecules \citep{onoa03}. The magnitude of the error increases as
the boundary conditions become more constrained, or when beads are
included in the model.

We emphasize that our fitting result is not directly comparable to the
fitting of experimental data, because our simulated ``data'' are not
noisy. In fits of real data, the quality of the fit is affected by the
amount of noise. However, our result does show the types of errors
that occur: fitting ``data'' generated by the FWLC to the usual WLC
solution leads to large errors in the persistence length for small
$L$. By contrast, fitting to the FWLC recovers the correct value of
$A$ .

\section*{Discussion}

In this paper we have developed the finite worm-like chain theory
(FWLC), which predicts the force-elasticity behavior of polymers with
short contour lengths ($\ell=L/A \sim $ 1--10). In addition to
retaining contributions from the subleading terms of equation
\eqref{znew}, the FWLC includes two
physical effects neglected in the usual WLC treatment: ({\it i})
chain-end boundary conditions and ({\it ii}) bead rotational
fluctuations.  Together, these effects may explain the apparent
decrease in polymer persistence length which has been experimentally
measured for short chains \citep{onoa03}.

A key result of this paper is the delineation of the regimes where we
expect the usual WLC solution to fail. We demonstrate that the FWLC
converges to the WLC in the limit of long contour length for all
forces, boundary conditions, and bead radii. However, for shorter
molecules the contour length crossover where the usual WLC solution
becomes inaccurate depends on applied force, chain-end boundary
conditions, and bead radius. For the FWLC and WLC to agree within 5\%
for a force-extension measurement with applied force of 0.1--10 pN,
requires $L/A > 100 $ ($L>5000$ nm for dsDNA).  

Although the FWLC improves on the usual WLC solution for short contour
lengths, it does not include all physical effects which occur in
single-molecule force microscopy. In particular, effects omitted from
the FWLC are important for very low or zero applied force (such as in
tethered particle motion experiments). These effects include the
chain-wall, chain-bead, and bead-wall exclusion interactions. All of
these interactions are most important at zero applied force. In our
calculations, the average position of the bead is ``outside'' the wall
for forces above 0.1 pN. For accurate predictions at lower applied
force, exclusion effects need to be included.  Comparison of the FWLC
with models which include the exclusion interactions can further
define the regimes of applicability of this model. In the future, an
improved theory which also applies at very low force may become
necessary.

The FWLC will be useful for experiments that use short NA polymers in
single-molecule force microscopy. A theory that is correct for small
$L/A$ helps ensure that the experimental force calibration is correct,
that least-squares fitting indeed recovers the correct values of the
contour and persistence lengths, and that the number of polymers
attached between surface and bead can accurately be determined.
The FWLC will facilitate experimental work with shorter polymers and
contribute to future discoveries in nucleic acid-enzyme interactions.

\section*{Acknowledgements}

We thank Igor Kulic, Tom Perkins, Rob Phillips, and Michael Woodside
for useful discussions, and the Aspen Center for Physics, where part
of this work was done. PCN acknowledges support from NSF grant
DMR-0404674 and the NSF-funded NSEC on Molecular Function at the
Nano/Bio Interface, DMR-0425780. MDB acknowledges support from NSF
NIRT grant PHY-0404286, the Butcher Foundation, and the Alfred
P. Sloan Foundation.



\clearpage 
\section*{Figure Legends}

\subsubsection*{Figure~\ref{expgeom}.}
Sketch of typical experimental geometries for single-molecule force
microscopy with nucleic acid polymers. The total molecule extension is
$z$. (a) One-bead geometry. The molecule is attached at one end to a
surface (glass slide or coverslip) and at other end to a bead. Force
is applied to the center of the bead. (b) Two-bead geometry. The
molecule is attached at both ends to beads. Force is effectively
applied to the centers of both beads.

\subsubsection*{Figure~\ref{coord}.}
Variables used to describe chain conformations.  Force is applied in
the $\hat{\bf z}$ direction. The arc length along the chain is $s$,
and the vector ${\bf r}(s)=(x(s),y(s),z(s))$ is the coordinate of the
chain at arc length $s$. The vector $\hat{{\bf t}}(s)$ is the unit
vector tangent to the chain at $s$.

\subsubsection*{Figure~\ref{beadrot}.}
Variables used to describe bead rotational fluctuations.  The vector
$\hat{\bf n}$ points from the center of the bead to the polymer-bead
attachment. The angle $\theta$ between the $\hat{\bf z}$ direction and
the bead-chain attachment is defined by $\cos \theta =-\hat{\bf n}
\cdot \hat{\bf z}$. The bead has radius $R$.

\subsubsection*{Figure~\ref{boundcond}.}
Sketch of the tangent-angle boundary conditions. (a) Unconstrained
boundary conditions. The tangent vector at the end of the chain (open
arrowhead) can point in all directions relative to the $\hat{\bf z}$ axis
(notched arrowhead) with equal probability. (b) Half-constrained
boundary conditions. Due to a constraint (such as the presence of a
solid surface normal to the $\hat{\bf z}$ axis) the tangent vector at
the end of the chain can point only in the upper half-sphere. (c) Perpendicular
boundary conditions. The tangent vector is constrained to point only
normal to the surface, in the $\hat{\bf z}$ direction.

\subsubsection*{Figure~\ref{beadbc}.}
Effective chain-end boundary conditions induced by bead rotational
fluctuations. The unnormalized probability density $g(\hat{{\bf t}}
\cdot \hat{\bf z})$ is shown versus $\cos \gamma = \hat{{\bf t}} \cdot
\hat{\bf z}$. Top, bead radius $R=100$ nm. Middle, bead radius $R=250$
nm. Bottom, bead radius $R=500$ nm. The different curves correspond to
different values of the applied force. In this calculation the
attachment of the chain to the bead is half-constrained: in the
absence of bead rotational fluctuations, the probability density is
zero for $\cos \gamma<0$ and constant for $\cos \gamma >0$. This step
function is approached for large force and large $R$. For smaller
values of $F R$ the probability density is smeared out, and approaches
a constant value (independent of $\cos \gamma$) for small $F R$.

\subsubsection*{Figure~\ref{infconv}.}
Convergence of the FWLC-predicted extension to the classic WLC
prediction for large $L$. The predicted fractional extension $z/L$ is
shown as a function of the contour length $L$ (at fixed applied
force). The different curves correspond to different theoretical
assumptions: ({\it i}) the classic WLC solution, calculated using the
method of ref.\  \citep{marko95}, ({\it ii}) the FWLC with unconstrained
boundary conditions, (\textit{iii}) the FWLC with half-constrained
boundary conditions , and (\textit{iv}) the FWLC with perpendicular boundary
conditions (see figure \ref{boundcond}).  The different panels show
different applied forces: top, 0.01 pN; middle, 0.1 pN; bottom, 1 pN.
The polymer persistence length is $A=50$ nm. Note that the WLC prediction
is independent of contour length, and the FWLC predictions converge to
the WLC prediction as $L$ increases. The convergence occurs more
quickly for higher applied force (see text, note different $y$-axis
scales for different panels). For low applied force, the more
constrained boundary conditions lead to larger predicted fractional
extension (see text).

\subsubsection*{Figure~\ref{errthresh}.}
Error threshold length $L_e$ as a function of force.  The error
threshold length is the contour length where the FWLC- and
WLC-predicted extensions differ by 5\%. For values of $L<L_e$, the
classic WLC solution gives significant errors in the predicted
extension. The polymer persistence length is $A=50$ nm. The error
threshold length is largest at low force. 

Top: no bead.  The different curves correspond to unconstrained,
half-constrained, and perpendicular boundary conditions.  The more
constrained boundary conditions have larger error threshold lengths at
low force than the unconstrained boundary condition. Comparable
predicted extensions by the classic WLC solution and the FWLC over the
range of forces and boundary conditions shown requires $L>10^4$ nm
($L/A>200$). 

Middle: one bead. The boundary conditions are half-constrained on both
ends of the chain. We calculate the extension predicted to the center
of the bead, then subtract the bead radius and compare to the
extension predicted by the WLC.  Larger bead radii lead to larger
values of $L_e$ at small forces, while for larger values of the
applied force $L_e$ is independent of the bead radius.  Comparable
predicted extensions by the classic WLC solution and the FWLC over the
range of forces and boundary conditions shown requires $L>8 \times
10^4$ nm ($L/A>1600$).

Bottom: two beads. The boundary conditions are half-constrained on
both ends of the chain. We calculate the extension predicted between
the centers of the beads, then subtract twice the bead radius and
compare to the extension predicted by the WLC. Larger bead radii lead
to larger values of $L_e$ at small forces, while for larger values of
the applied force, $L_e$ is independent of the bead radius.  The
values of $L_e$ are larger for the two-bead case than the one-bead
case.  Comparable predicted extensions by the classic WLC solution and
the FWLC over the range of forces and boundary conditions shown
requires $L> 10^5$ nm ($L/A>2000$).

\subsubsection*{Figure~\ref{forcextfig1}.}
Force-extension curves predicted by the FWLC model for $L=200$ nm
($L/A=4$). 

Top: no bead. The different curves correspond to the classic WLC
solution and the FWLC with unconstrained, half-constrained, and perpendicular
boundary conditions. The predicted extension for all boundary
conditions converges at high force. For low applied force, the more
constrained boundary conditions lead to larger predicted extension
(see text).

Middle: one bead. The different curves correspond to bead radii of
100, 250, and 500 nm. The predicted molecule extension (after the bead
radius has been subtracted) agrees with the no-bead prediction for
large force, but decreases more rapidly as the force decreases. For
low applied force (below 0.05 pN) the predicted extension becomes
negative, because the bead-wall exclusion interaction is neglected in
the model. In this regime the theory gives unphysical predictions.

Bottom: two beads. The different curves correspond to bead radii of
100, 250, and 500 nm. The predicted molecule extension (after twice
the bead radius has been subtracted) agrees with the no-bead and
one-bead predictions for large force. For low applied force (below
0.08 pN) the predicted extension becomes negative, because the
bead-bead exclusion interaction is neglected in the model. In this
regime the theory gives unphysical predictions.

\subsubsection*{Figure~\ref{forcextfig2}.}
Force-extension curves predicted by the FWLC model for $L=500$ nm
($L/A=10$). 
 Compared to the curves for $L=200$ nm (figure
\ref{forcextfig1}) the effects of boundary conditions and beads are
less important.

Top: no bead. The different curves correspond to the classic WLC
solution and the FWLC with unconstrained, half-constrained, and perpendicular
boundary conditions. 

Middle: one bead. The different curves correspond to bead radii of
100, 250, and 500 nm.
For
low applied force (below 0.02 pN) the predicted extension becomes
negative, because the bead-wall exclusion interaction is neglected in
the model. In this regime the theory gives unphysical predictions.

Bottom: two beads. The different curves correspond to bead radii of
100, 250, and 500 nm. 
For low applied force (below
0.06 pN) the predicted extension becomes negative, because the
bead-bead exclusion interaction is neglected in the model. In this
regime the theory gives unphysical predictions.

\begin{figure}
   \begin{center}
      \includegraphics*[width=3in]{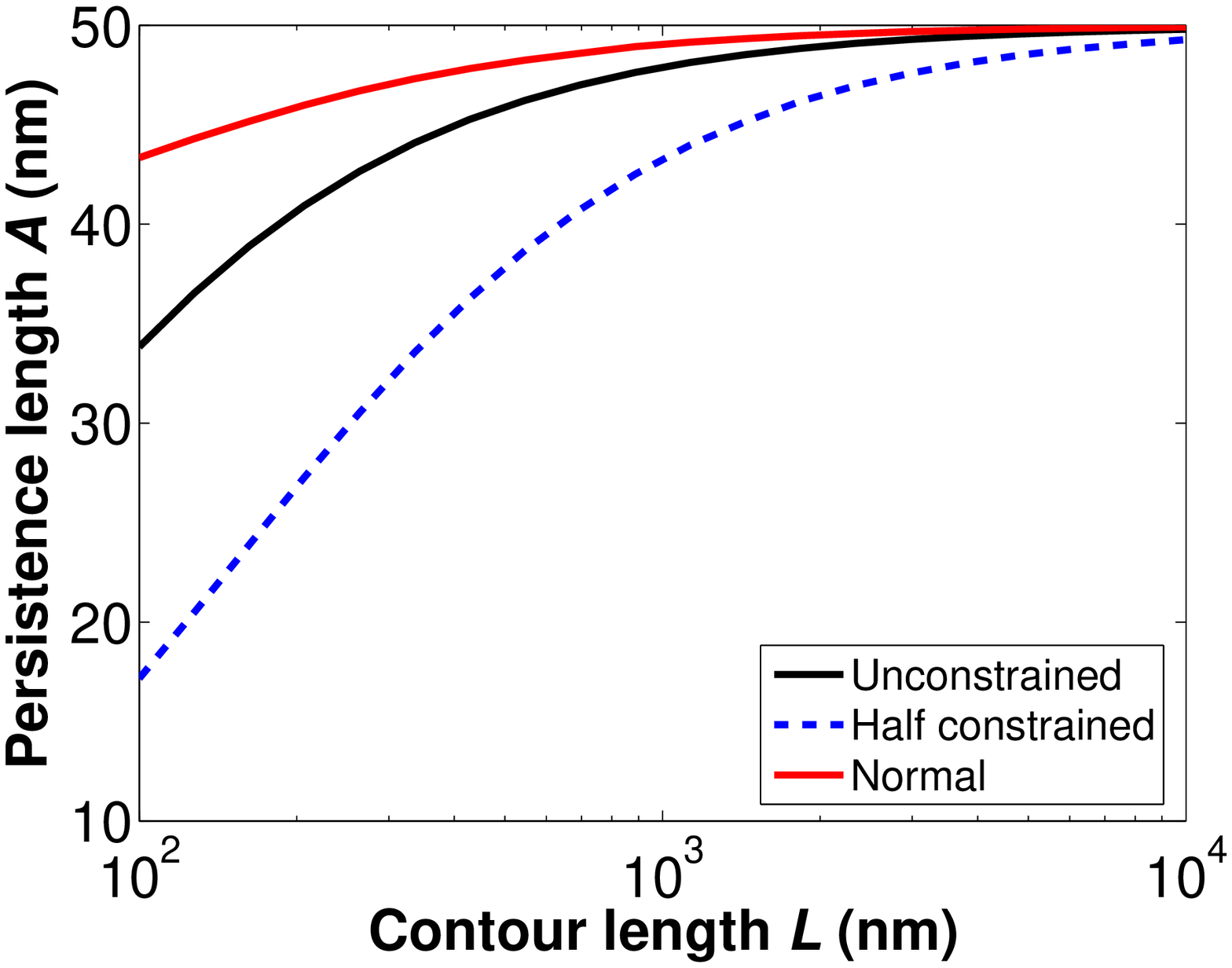}\\
      \includegraphics*[width=3in]{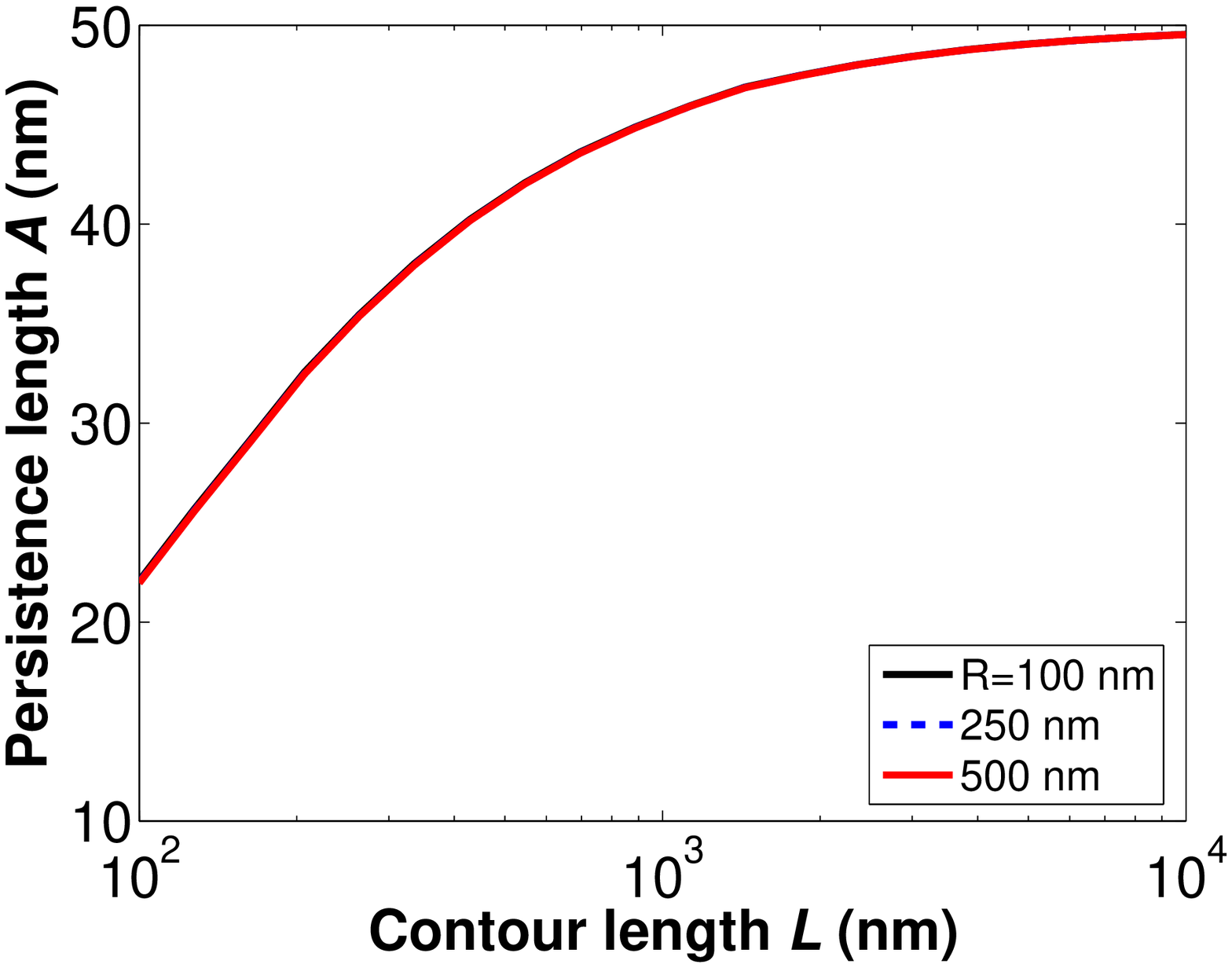}\\
      \includegraphics*[width=3in]{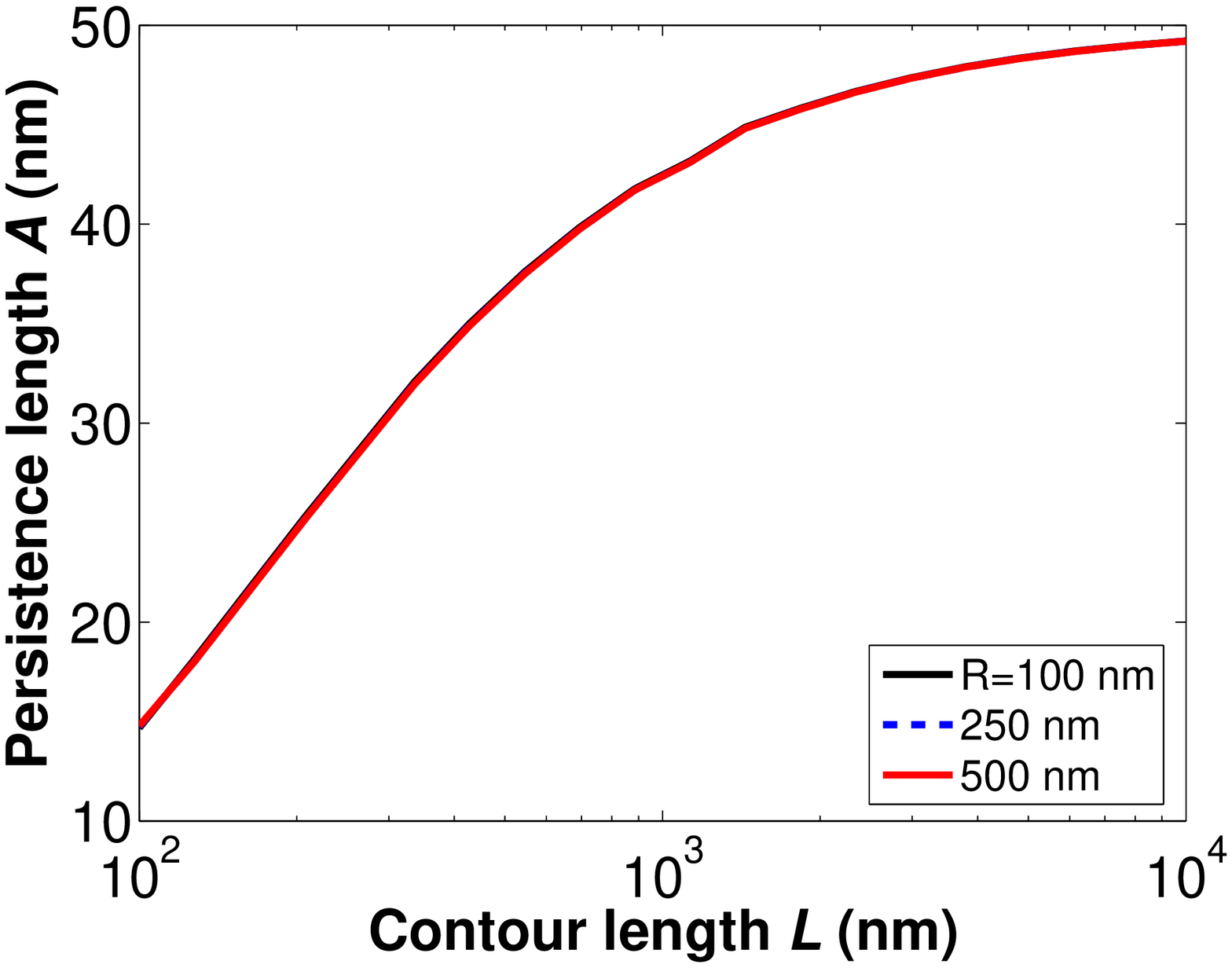}
      \caption{}
      \label{fitper}
   \end{center}
\end{figure}

\subsubsection*{Figure~\ref{fitper}.}
Results of fitting the FWLC model predictions to the classic WLC
solution. The fit value of the persistence length is shown as a
function of contour length $L$.  The polymer persistence length is
$A=50$ nm.

Top: no bead. The different curves correspond to the unconstrained,
half-constrained, and perpendicular boundary conditions. The apparent $A$
decreases most for the half-constrained boundary conditions. The
apparent persistence length decreases by more than a factor of two as
$L$ decreases from $10^4$ to 100 nm.

Middle: one bead. The different curves correspond to bead radii of
100, 250, and 500 nm. The decrease in the apparent persistence length
is similar to that shown in the top panel, half-constrained boundary
conditions.
 
Bottom: two beads. The different curves correspond to bead radii of
100, 250, and 500 nm. The apparent persistence length decreases more
quickly with contour length than for the one-bead case.

\clearpage  

\clearpage

\clearpage

\clearpage  

\clearpage  

\clearpage  

\clearpage  

\clearpage  
\clearpage  

\clearpage  

\end{document}